\input harvmac
\input epsf

\newcount\figno
\figno=0
\def\fig#1#2#3{
\par\begingroup\parindent=0pt\leftskip=1cm\rightskip=1cm\parindent=0pt
\baselineskip=12pt
\global\advance\figno by 1
\midinsert
\epsfxsize=#3
\centerline{\epsfbox{#2}}
\vskip 14pt

{\bf Fig. \the\figno:} #1\par
\endinsert\endgroup\par
}
\def\figlabel#1{\xdef#1{\the\figno}}
\def\encadremath#1{\vbox{\hrule\hbox{\vrule\kern8pt\vbox{\kern8pt
\hbox{$\displaystyle #1$}\kern8pt}
\kern8pt\vrule}\hrule}}

\overfullrule=0pt

\noblackbox
\parskip=1.5mm

\def\Title#1#2{\rightline{#1}\ifx\answ\bigans\nopagenumbers\pageno0
\else\pageno1\vskip.5in\fi \centerline{\titlefont #2}\vskip .3in}

\font\caps=cmcsc10

\noblackbox
\parskip=1.5mm

  
\def\npb#1#2#3{{\it Nucl. Phys.} {\bf B#1} (#2) #3 }
\def\plb#1#2#3{{ \it Phys. Lett.} {\bf B#1} (#2) #3 }
\def\prd#1#2#3{{ \it Phys. Rev. } {\bf D#1} (#2) #3 }
\def\prl#1#2#3{{ \it Phys. Rev. Lett.} {\bf #1} (#2) #3 }

\def\ijmpa#1#2#3{{ \it Int. J. Mod. Phys.} {\bf A#1} (#2) #3 }
\def\jmp#1#2#3{{ \it J. Math. Phys.} {\bf #1} (#2) #3 }
\def\cmp#1#2#3{{\it Commun. Math. Phys.} {\bf #1} (#2) #3 }

\def\bb#1{{\tt hep-th/#1}}

\def\mathph#1{{\tt math-ph/#1}}

\def\jhep#1#2#3{{ \it J. High Energy Phys.} {\bf #1} (#2) #3 }


           \def\CO{{\cal O}} 
\def\CA{{\cal A}} \def\CC{{\cal C}}  
\def\CL{{\cal L}} \def\CH{{\cal H}}  
   
\def\CM{{\cal M}} 
\def\CN{{\cal N}}


\def\dj{\hbox{d\kern-0.347em \vrule width 0.3em height 1.252ex depth
-1.21ex \kern 0.051em}}

\def\half{{1\over 2}\,}

\def\Tr{{\rm Tr\,}}
\def\tr{{\rm tr\,}}

\def\ubar{\overline{U(N)}_\Theta} 
\def\Dirac{\,\raise.15ex\hbox{/}\mkern-13.5mu D}
\def\dirac{\,\raise.15ex\hbox{/}\kern-.57em \partial}
\def\shalf{{\ifinner {\textstyle {1 \over 2}}\else {1 \over 2} \fi}} 
\def\sshalf{{\ifinner {\scriptstyle {1 \over 2}}\else {1 \over 2} \fi}} 
\def\sfourth{{\ifinner {\textstyle {1 \over 4}}\else {1 \over 4} \fi}}
\lref\rrev{M.R. Douglas and N.A. Nekrasov, \bb{0106148\semi} 
A. Konechny and A. Schwarz, \bb{0012145\semi}
R.J. Szabo, \bb{0109162.}}

\lref\rbasic{A. Connes, {\it ``Noncommutative Geometry"}, Academic Press
1994.} 

\lref\rrotr{J. Madore, {\it ``An Introduction to Noncommutative Geometry
and its Physical Applications,"} Cambridge Univ. Press. 1999 (LMS Lecture Notes
Series 257, 2nd ed.)\semi
J.M. Gracia-Bondia, J.C. Varilly and H. Figueroa, {\it ``Elements of Noncommutative Geometry,"} Birkhaeuser 2001, Boston, USA.} 

\lref\rdipoles{C.-S. Chu
and P.-M. Ho,
\npb{550}{1999}{151} \bb{9812219\semi}
 M.M. Sheikh-Jabbari, \plb{455}{1999}{129} \bb{9901080\semi}
D. Bigatti and L. Susskind, \prd{62}{2000}{066004}  \bb{9908056.} 
}
\lref\rnust{N. Seiberg, L. Susskind and N. Toumbas, \jhep{06}{2000}{044}
\bb{0005015,} \jhep{06}{2000}{021} \bb{0005040\semi} R. Gopakumar, J. Maldacena,
J. Minwalla and S. Strominger, \jhep{06}{2000}{036} \bb{0005048\semi} J.L.F.
Barb\'on and E. Rabinovici, \plb{486}{2000}{202} \bb{0005073.}}

\lref\rnuft{ J. Gomis and T. Mehen, \npb{591}{2000}{265} \bb{0005129\semi}
L. Alvarez-Gaum\'e, J.L.F. Barb\'on and R. Zwicky, \jhep{05}{2001}{057}
\bb{0103069.}}

\lref\ruvir{S. Minwalla, M. Van Raamsdonk and N. Seiberg, \jhep{02}{2000}{020}
\bb{9912072.}} 

\lref\rmorita{A. Schwarz, \npb{534}{1998}{720} \bb{9805034\semi} M.A. Rieffel
and A. Schwarz, {\tt math.QA/9803057\semi} M.A. Rieffel, {\it
J. Diff. Geom.} {\bf 31}
(1998) 535 {\tt quant-ph/9712009.}}

\lref\ritalia{G. Landi, F. Lizzi and R.J. Szabo, 
\cmp{206}{1999}{603} 
\bb{9806099.}}

\lref\rmortd{D. Brace, B. Morariu and B. Zumino, \npb{545}{1999}{192} 
\bb{9810099,} \npb{549}{1999}{181} \bb{9811213\semi}
 C. Hofman and E. Verlinde,
\jhep{12}{1998}{010} \bb{9810116,} \npb{547}{1999}{157} \bb{9810219\semi}
B. Pioline and A. Schwarz, \jhep{08}{1999}{021} \bb{9908019.}}

\lref\rhover{ C. Hofman and E. Verlinde,
 \npb{547}{1999}{157} \bb{9810219.}} 

\lref\rzack{Z. Guralnik and J. Troost, \jhep{05}{2001}{022}  \bb{0103168.}}

\lref\rzackul{Z. Guralnik, \bb{0109079.}} 
 
\lref\rtoni{A. Gonz\'alez-Arroyo and C.P. Korthals-Altes, \plb{131}{1983}{396.}
}

\lref\rtek{A. Gonz\'alez-Arroyo and M. Okawa, \prd{27}{1983}{2397\semi} T. Eguchi
and R. Nakayama, \plb{122}{1983}{59.}}

\lref\rnbi{J. Ambjorn, Y.M. Makeenko, J. Nishimura and R.J. Szabo,
\jhep{05}{2000}{023} \bb{0004147.}}

\lref\rmat{G. Landi, F. Lizzi and R.J. Szabo, 
\cmp{217}{2001}{181} 
\bb{9912130.}}

\lref\rfilk{T. Filk, \plb{376}{1996}{53.}}

\lref\rpepeg{J.C. V\'arilly and J.M. Gracia-Bond\'{\i}a, \ijmpa{14}{1999}{1305}
\bb{9804001.}} 

\lref\rafree{
C.P. Mart\'{\i}n and D. S\'anchez-Ruiz, \prl{83}{1999}{476} \bb{9903077\semi}
M.M. Sheikh-Jabbari, \jhep{06}{1999}{015} \bb{9903107\semi}
T. Krajewski and R. Wulkenhaar, \ijmpa{15}{2000}{1011}  
 \bb{9903187.}} 

\lref\rruso{K. Saraikin, {\it J. Exp. Theor. Phys.} {\bf 91} 2000 653  
\bb{0005138.}} 

\lref\rbps{A. Konechny and A. Schwarz, \npb{550}{1999}{561} \bb{9811159,}
\plb{453}{1999}{23} \bb{9901077,} \jhep{09}{1999}{030} \bb{9907008.}}

\lref\rirr{S. Elitzur, B. Pioline and E. Rabinovici, \jhep{10}{2000}{011}
\bb{0009009.}} 

\lref\rbook{H.E. Rose, {\it ``A Course in Number Theory"}, Oxford University  
Press (1994).} 

\lref\rHI{A. Hashimoto and N. Itzhaki, \jhep{12}{1999}{007} \bb{9911057.}}

\lref\rlargen{G. 't Hooft, \npb{75}{1974}{461.}} 

\lref\rwulk{T. Krajewski and R. Wulkenhaar, \ijmpa{15}{2000}{1011}
 \bb{9903187.}}

\lref\rdurham{V.V. Khoze and G. Travaglini, \jhep{01}{2001}{026} 
\bb{0011218.}} 

\lref\rfernando{C.P. Mart\'{\i}n and F. Ruiz-Ruiz, \npb{597}{2001}{197}
\bb{0007131\semi} F. Ruiz-Ruiz, \plb{502}{2001}{274} \bb{0012171.}}

\lref\rsut{
M. Hayakawa, 
\plb{478}{2000}{394} \bb{9912094,} 
 \bb{9912167\semi} 
A. Matusis, L. Susskind and N. Toumbas, \jhep{12}{2000}{002}   
\bb{0002075\semi}
I.Ya. Aref'eva, D.M. Belov and A.S. Koshelev, \bb{0003176.}} 

\lref\rgomwise{J. Gomis, T. Mehen and M.B. Wise, \jhep{08}{2000}{029}
  \bb{0006160.}} 

\lref\respe{K. Landsteiner, E. L\'opez and M.H.G. Tytgat, 
\jhep{09}{2000}{027} \bb{0006210,} \jhep{06}{2001}{055} \bb{0104133.}} 

\lref\rchucs{C.-S. Chu, \npb{580}{2000}{352}  \bb{0003007.}}

\lref\rshift{G.-H. Chen and Y.-S. Wu, \npb{593}{2001}{562} \bb{0006114.}} 
 
\lref\rcarmelo{A. Das and M.M. Sheikh-Jabbari, \jhep{06}{2001}{028}
\bb{0103139\semi} C.P. Mart\'{\i}n, \plb{515}{2001}{185}  \bb{0104091.}}
 
\lref\rgubser{S.S. Gubser and S.L. Sondhi, \npb{605}{2001}{395}
 \bb{0006119.}}

\lref\rrabisan{J.L.F. Barb\'on and E. Rabinovici, 
\jhep{12}{1999}{017} \bb{9910019.}}

\lref\rwitclass{E. Witten, \npb{202}{1982}{253.}}

\lref\rwittd{E. Witten, \bb{9903005.}}

\lref\rwitfd{E. Witten, \bb{0006010.}}

\lref\rthooft{G. 't Hooft, \npb{138}{1978}{1,} \npb{153}{1979}{141,}
\npb{205}{1982}{1.}}

\lref\rrevbaal{P. van Baal, {\tt hep-ph/0008206}.}  

\lref\rrevtoni{A. Gonz\'alez-Arroyo, \bb{9807108.}} 

\lref\rusmat{J.L.F. Barb\'on and M. Garc\'{\i}a P\'erez, to appear.}

\lref\rshif{A. Kovner and M. Shifman, \prd{56}{1997}{2396} \bb{9611213.}} 

\lref\rncinst{N. Nekrasov and A. Schwarz, \cmp{198}{1998}{689}
 $\,$\bb{9802068\semi}  A. Astashkevich,
N. Nekrasov and A. Schwarz, \cmp{211}{2000}{167} \bb{9810147.}}

\lref\rharvey{J.A. Harvey, \bb{0105242.}}

\lref\rszabo{F. Lizzi, R.J. Szabo and A. Zampini, 
\jhep{08}{2001}{032} 
\bb{0107115.}} 

\lref\rpalais{R.S. Palais, {\it Topology} {\bf 3} (1965) 271.}

\lref\robliq{G. 't Hooft, {\it ``The Confinement Phenomenon in Quantum
Field Theory"}, Carg\`ese Summer School Lecture Notes on Fundamental
Interactions, Eds. M L\'evy and J.-L. Basdevant, {\it Nato Adv.
Study Inst. Series {\bf B}: Phys.} Vol 85 (1981) 639.} 

\lref\rwitdyon{E. Witten, \plb{86}{1979}{283.}} 

\lref\rwitjons{E. Witten, \cmp{121}{1989}{351.}}

\lref\rmargas{M. Garc\'{\i}a-P\'erez and A. Gonz\'alez-Arroyo,   
{\it J. Phys} {\bf A} 26 (1993) 2667 \semi  
A. Montero, \jhep{0005}{2000}{022}  
{\tt hep-lat/0004009.}}  

\lref\rholo{A. Pinzul and A. Stern, \bb{0107179.}}
 
\lref\rpoly{A.P. Polychronakos,  
{\it Annals of Physics} {\bf 203} (1990) 231.} 

\lref\rquank{V.P. Nair and A.P. Polychronakos, \prl{87}{2001}{030403}  
 \bb{0102181\semi} D. Bak,
K. Lee and J. Park, \prl{87}{2001}{030402}  
\bb{0102188\semi} 
M.M. Sheikh-Jabbari, \plb{510}{2001}{247} \bb{0102092.}} 

\lref\rkraj{T. Krajewski, {\tt math-ph/9810015.}} 

\lref\rsuskqh{L. Susskind, \bb{0101029.}}

\lref\rbqh{A.P. Polychronakos, \jhep{04}{2001}{011}  
 \bb{0103013,} \jhep{06}{2001}{070} 
\bb{0106011.} }   

\lref\rzacku{Z. Guralnik, \bb{9804057.}} 

\lref\rgrandi{N. Grandi and G.A. Silva, \plb{507}{2001}{345}  \bb{0010113.}} 

\lref\rram{S. Hellerman and  M. Van Raamsdonk,  
\bb{0103179.}}
 
\lref\rcsqh{S.C. Zhang, T.H. Hansson and S. Kivelson, \prl{62}{1988}{82.}}

\lref\rbacsus{S. Bahcall and L. Susskind, {\it Int. J. Mod. Phys.} {\bf B5}
(1991) 2735.}

\lref\rfzee{J. Frohlich and A. Zee, \npb{364}{1991}{517.}}

\lref\rfroh{J.  Frohlich, B.  Pedrini, C. Schweigert and  J. Walcher,
{\tt cond-mat/0002330.}} 
 
\lref\rwen{X.-G. Wen, {\it Phys. Rev.} {\bf B40} (1989) 7387, {\it
Int. J. Mod. Phys.} {\bf B4} (1990) 239, {\it Int. J. Mod. Phys.} {\bf B5}
(1991) 1641.}

\lref\rcapp{A. Cappelli, C.A. Trugenberger and G.R. Zemba, \npb{448}{1995}{470}
\bb{9502021.}} 

\lref\rcappelli{A. Cappelli, C.A. Trugenberger and G.R. Zemba, {\bb 9610019.}}

\lref\rzac{D.B. Fairlie, P. Fletcher and C.K. Zachos, \jmp{31}{1990}{1088.}}

\lref\mac{A.J. Macfarlane, A. Sudbery and P.H. Weisz, \cmp{11}{1968}{77.}}

\lref\rarmoni{A. Armoni, \npb{593}{2001}{229}  
\bb{0005208.}} 
 
\lref\kphd{T. Krajewski, {\it G\'eom\'etrie non commutative et interactions
fondamaentales}, Ph.D. Thesis. (\mathph{9903047})}

\lref\rSW{N. Seiberg and E. Witten, \jhep{09}{1999}{032,} \bb{9908142.}}

\lref\rncads{J.M. Maldacena and J.G. Russo, \jhep{09}{1999}{25} \bb{9908134\semi
}
A. Hashimoto and N. Itzhaki, \plb{465}{1999}{142} \bb{9907166.}}

\lref\rcds{A. Connes, M.R. Douglas and A. Schwarz, \jhep{02}{1998}{003,} 
\bb{9711162.} M.R. Douglas and C.M. Hull, \jhep{02}{1998}{008,} \bb{9711165.}
 }

\lref\rreview{M.R. Douglas and N.A. Nekrasov, 
\bb{0106048.}}


\baselineskip=15pt

\line{\hfill CERN-TH/2001-240}
\line{\hfill {\tt hep-th/0109176}}

\vskip 0.7cm

\Title{\vbox{\baselineskip 12pt\hbox{}
 }}
{\vbox {\centerline{Morita Duality and Large-$N$ Limits }
}}

\vskip0.7cm

\centerline{$\quad$ {\caps L. Alvarez-Gaum\'e and 
J.L.F. Barb\'on\foot{ On leave
from Departamento de F\'{\i}sica de Part\'{\i}culas da
 Universidade de Santiago de Compostela, Spain.} }} 
\vskip0.7cm

\centerline{{\sl  Theory Division, CERN, 
 CH-1211 Geneva 23, Switzerland}}
\centerline{{\tt luis.alvarez-gaume@cern.ch,
barbon@mail.cern.ch}} 

\vskip1cm

\centerline{\bf ABSTRACT}

 \vskip 0.1cm

 \noindent 
We study some dynamical aspects of gauge theories on noncommutative
tori. We show that Morita duality, combined with the hypothesis of
analyticity as a function of the noncommutativity parameter $\Theta$, gives
information about singular large-$N$ limits of ordinary $U(N)$ gauge
theories, where the large-rank limit is correlated with the shrinking
of a two-torus to zero size.  We study some non-perturbative  tests of
the smoothness hypothesis
with respect to $\Theta$  in theories
with and without supersymmetry.  In the supersymmetric case this is
done by adapting Witten's index to the present situation, and in the
nonsupersymmetric case by studying the dependence of energy levels
on the instanton angle.  We find that regularizations which restore
supersymmetry at high energies seem to preserve $\Theta$-smoothness
whereas nonsupersymmetric asymptotically free theories seem to violate
it.  As a final application we use Morita duality to study a recent
proposal of Susskind to use a noncommutative Chern-Simons gauge theory
as an effective description of the Fractional Hall Effect.  In particular
we obtain an elegant derivation of Wen's topological order.

\vskip 0.1cm

\Date{September 2001}
               

\vfill





\baselineskip=14pt

\newsec{Introduction}

\noindent

Noncommutative Field Theories (NCFTs) provide an interesting generalization
of the framework of local quantum field theory, allowing for some degree
of non-locality while still keeping an interesting mathematical structure
\refs{\rbasic, \rrotr, \rrev}.
From a different point of view,
 the perturbative dynamics of NCFT mimics in many respects
that of string theory. To be more specific,
NCFT arises as a peculiar low-energy limit of open-string theory,
\refs{\rcds,\rSW}
 so that the string becomes a rigid, extended
object, i.e. a rigid `dipole' \refs\rdipoles. This is achieved by placing the
strings in a large magnetic field. The resulting low-energy
theory lives effectively   on a noncommutative  version
of ${\bf R}^d$, where the non-locality is parametrized by a quantum phase-space
structure of  space-time:
$
[x^\mu, x^\nu] =i\, \theta^{\mu\nu}$,
with $\theta^{\mu\nu}$  related to the string magnetic field $B_{\mu\nu}$
via $\theta = B^{-1}$. The stringy nature of NCFT is not really hidden in
the short-distance structure, but apparent at low energies, since the
size of the `dipoles' is approximately linear in the momentum:
 $\ell_{\rm eff} = |\theta^{\mu\nu} p_\nu | $, so
that high energy particles are macroscopic in size. This is the basis for
the so-called UV/IR connection, \refs{\ruvir} that represents
the main novelty of NCFT dynamics. In particular, if time enters the
noncommutativity relations, the particles are non-local in time, which
poses a problem for Hamiltonian methods. In fact, these  theories
appear to be inconsistent when defined as field theories with a finite
number of particle degrees of freedom \refs{\rnust, \rnuft}. We henceforth
restric the noncommutativity of space--time to the purely spatial sections.

Another `stringy' property of NCFT is  the so-called Morita duality of
gauge theories on noncommutative tori  \refs{\rbasic,\rcds, \rmorita}
  a low-energy remnant of the T-duality
symmetry of the underlying string model \refs\rmortd\ (see also \refs\ritalia.)
 It acts on
the periods of the string magnetic field
 $\int B/ 2\pi$ by fractional linear
transformations. In particular, if the  matrix of periods has rational entries,
there is a Morita transformation that maps the given noncommutative theory to
an ordinary $U(N)$ gauge theory on a smaller torus, with a larger rank
and some non-vanishing magnetic fluxes. This opens up the
interesting possibility of using the information encoded in the
$\theta$-dependence of physical quantities to learn about ordinary
gauge theories at finite volume, i.e. the complicated structure of
confining, oblique-confining, Higgs and Coulomb phases of gauge theories
could be encoded in the noncommutative language in terms of some interesting
modular properties as a function of $\theta$.
 Conversely,  the exotic features of
perturbation theory in NCFT, notably the UV/IR phenomenon,
 put into question most of our  standard expectations
for non-perturbative dynamics in NCFT. In this context, Morita duality
can be used to translate the standard body of knowledge about ordinary
 confining
gauge theories into the context of their noncommutative cousins.
In particular, it was suggested in \refs\rzack\ that smooth behaviour
of physical quantities in $\theta$  would imply very nontrivial constraints
on the large-$N$ limit of ordinary gauge theories on tori.

The interplay between noncommutativity and the
 large-$N$ limit is at the root
of the subject in its (hidden) beginnings \refs\rtoni, since NCFT
   is   a formal  continuum limit of  the twisted reduced Eguchi--Kawai
models \refs\rtek\ (see also \refs{\rnbi, \rmat}).
In this article we sharpen this relationship with a number of
qualitative and quantitative
 tests. We  propose an analiticity criterion on $\theta$-dependence
based on the limiting behaviour of rational approximations to generic
NCFTs.  We show that this criterion of $\theta$-smoothness gives information
about ordinary gauge theories in a {\it singular} large-$N$ limit.
Although this singular limit is difficult to study in the language of
ordinary gauge theory, we argue that its dynamics can be very rich.

We also add some comments on the inverse
problem, i.e. we study degenerate limits of NCFTs which are tuned to
reproduce the standard confinement regime of ordinary gauge theories
in the large-$N$ limit.  Morita duality implies that
 all these limits necessarily involve infinite noncommutativity $\theta
\rightarrow \infty$, a property
 already hinted at by the behaviour of perturbation
theory \refs{\rtek, \rfilk, \ruvir}.

Some exact information about the singular large-rank
 limits is obtained by
 considering rational approximations of $\CN=1$ theories in three and
four dimensions. We define an appropriate Witten index that probes
aspects of confinement dynamics in    the non-abelian sector of the
models,  and show that this index varies smoothly
with $\theta$. In some cases this behaviour depends on subtle properties
such as oblique confinement in ordinary gauge theories.

In section five we go one step further by considering the dependence
of energy levels on the instanton angle in non-supersymmetric theories. For
a specific choice of electric  and magnetic fluxes,
a reliable estimate is possible within an instanton-gas approximation, since
the corresponding energy splittings receive no contribution in perturbation
theory.   This is a rather sensitive
test of $\Theta$-smoothness, because of the occurrence of level-crossing
phenomena that depend explicitly on the rank of the gauge theory.
We find that  a regularization with
 restored $\CN=4$  supersymmetry at high energies
 preserves $\Theta$-smoothness
of the low-energy physics,  whereas asymptotically free non-supersymmetric
theories seem to violate the smoothness constraints. Hence, some form
of supersymmetry, albeit `softly broken' seems to be necessary to
maintain continuity of the physics as a function of $\Theta$.

 We end with an application of Morita duality
to the Quantum Hall Effect by
calculating Wen's topological order \refs\rwen\ for the proposed
noncommutative Chern--Simons effective description of the Fractional
Hall Effect, \refs\rsuskqh.

\newsec{Morita Equivalence of Gauge Theories}

\noindent

We consider
 $U(N)$  Noncommutative Yang--Mills Theories (NCYM)
on ${\bf S}^1_\beta \times {\bf T}_\theta^{3}$,
 with ${\bf S}^1_\beta$ representing a compact euclidean time direction
 of length $\beta$, and ${\bf T}^3_\theta$ a noncommutative three-torus
with flat metric.
The microscopic parameters of the
$U(N)$ gauge theory  are
the Yang--Mills  coupling at some cutoff scale, $g$, the instanton
angle, $\vartheta$,
 and  the spatial noncommutativity
 parameters $\theta^{jk}$.
 There is also a   background field
$\phi_{ij}$ that enters the physics as a constant $U(1)$ shift
of the field strength, so that  the  action reads
\eqn\ac
{
S =     {1 \over 2g^2}\int
\tr\,(F+\phi)_{\mu\nu}  (F+\phi)^{\mu\nu}
+ {i\vartheta \over 8\pi^2} \int \tr\,(F+\phi) \wedge (F+\phi) + \dots,}
where
the dots stand for other terms depending on extra fields, such
as fermions or scalars in supersymmetric theories, all of them in
the adjoint representation of the gauge group, and $F$ is the noncommutative
gauge field strength.

The Feynman rules of this theory are obtained from those of the ordinary
$U(N)$ theory by the following replacement of the  $U(N)$ structure constants:
\eqn\refpl
{f^{\alpha\beta\gamma} \rightarrow f^{\alpha\beta\gamma}
 \,{\rm cos}\left(\shalf
\theta^{ij} k_i
k'_j\right) + d^{\alpha\beta\gamma}\,{\rm sin}\left(\shalf \theta^{ij} k_i k'_j
\right).
} 
$k,k'$ are any two momenta entering the trilinear vertex.
In particular, we see that the
global $U(1)$ is self-coupled and
coupled to the $SU(N)$ subgroup. Indeed, the rank-one
 $U(1)$  theory  shows asymptotic freedom \refs\rafree.

 It is useful to define  a dimensionless noncommutativity parameter and
background field by
\eqn\dimt
{\Theta^{ij} = {2\pi \over L_i L_j} \,\theta^{ij}, \qquad
\Phi_{ij} =
 {L_i L_j  \over 2\pi} \,\phi_{ij}.}

When considering the Hamiltonian quantization   on ${\bf T}^3$,
 the Hilbert space splits into
sectors labelled by       the integer magnetic fluxes
\eqn\magf{
m_{ij} =
\int_{(ij)}\; \tr\;{F \over 2\pi},}
 determined
by the first Chern class of the $U(N)$ gauge bundle,
together with the usual  integer
momenta 
\eqn\momf{
p_j = -i \,L_j \,\int_{{\bf T}_\theta^3} \tr\,{F_{jk} \over 2\pi}
 \, {\delta \over
\delta A_k}.}
There are also integrally quantized electric fluxes $w^j$ of the form
 (c.f. \refs\rhover):   
\eqn\elef{
w^j = -{i\over L_j} \int_{{\bf T}_\theta^3}
 \tr\, {\delta \over \delta A_j} - \Theta^{jk}\,p_k.}
We shall frecuently use the vector notation for the magnetic fluxes
and the  background field:
$$
m^k = \shalf\,\epsilon^{kij} \,m_{ij}, \qquad \Phi^k = \shalf\,
\epsilon^{kij} \,\Phi_{ij}.
$$

 For simplicity, we consider orthogonal three-tori  of the form ${\bf T}^3 =
 {\bf T}_\theta^2 \times {\bf S}_e^1$ with
 $\Theta$ and $\Phi$  of rank two and  aligned
with  the noncommutative two-torus ${\bf T}_\theta^2$,
 which will be assumed squared
with side length $L$. The length of ${\bf S}^1_e$ is
 $L_e$, so that the volume of the
 three-torus is $V=L_e L^2$.
 In this simple case  Morita duality is represented
 by the $SL(2,{\bf Z})$ action:
\eqn\mordu{
\eqalign{
\Theta' &= {b\Theta -a \over s+r\Theta}, \qquad \;\;\;\;\;
\Phi' = (s+r\Theta)^2 \Phi -
r(s+r\Theta), \cr
 \tau' &=
 {\tau \over |s+r\Theta|}, \qquad \;\;\; L' = |s+r\Theta|\,L
}}
on the parameters, where $a, b, s, r \in {\bf Z}$ and $sb+ar=1$.
We have collected the gauge coupling and instanton angle into the
complex coupling $2\pi\tau = \vartheta + 8\pi^2 i/g^2$.
The action on the other quantum numbers is
\eqn\adis{
 \left(\matrix{ m'
 \cr N' \cr}
\right) =
 \left(\matrix{ s & r
 \cr -a & b \cr}
\right)
 \left(\matrix{ m
 \cr N \cr}
\right), \qquad
 \left(\matrix{ p'
 \cr *w' \cr}
\right) =
 \left(\matrix{ s & r
 \cr -a & b \cr}
\right)
 \left(\matrix{ p
 \cr *w \cr}
\right)
,}
where $m$ is the magnetic flux through ${\bf T}^2_\theta$ and $ w, p$
 are the two-dimensional vectors of  electric fluxes and
momenta along the same torus. The notation $*w$ refers to the Hodge
duality operation on the ${\bf T}^2_\theta$, i.e. $(*w)_i = \epsilon_{ij} w^j$.

 In perturbation theory, Morita
equivalence  follows as an algebraic
identity of the Feynman diagram expansion
 \refs\rtek, or as T-duality of the regularized version in terms
of an underlying string model \refs\rmortd.
 More generally,
 there is a very explicit proof
 as a formal change of variables in a path integral \refs\rnbi.

Rational theories are characterized by a rational dimensionless
 noncommutativity
parameter.
If  $\Theta = a/b$, so that $s+r\Theta= 1/b$,
the Morita transformation \mordu\ sends
the theory  to an ordinary one with $\Theta' =0$ and different  parameters,
involving in particular a rescaling of the rank of the gauge group.
In this case, the details of the transformation at the level of fields
are rather elementary (c.f. \refs{\rzack,\rruso}).
 Consider, for example, a $U(N)$-valued
 noncommutative  connection on ${\bf T}^2_\theta \times Y$,
 which is also {\it periodic}
on the two-torus ${\bf T}^2_\theta$ of length $L$, and has arbitrary
boundary conditions on the commutative space $Y$:
\eqn\peri{
A(x,y)= \sum_{\ell \in {\bf Z}^2} a_\ell (y) \;e^{-2\pi i \ell \cdot x /L}.}
The coefficient matrices $a_\ell$ satisfy $a_\ell^\dagger = a_{-\ell}$
and have arbitrary dependence on the commutative coordinates $y\in Y$.

The  Morita-dual is a
 twisted $U(N')$ theory with $N' =N\,b$,  with a connection  given by
a particular superposition of matrices in the subgroup $U(N)\otimes U(b)$:
\eqn\ord{
A'(x,y) +A_{\phi'} (x,y) =
 \sum_{\ell\in {\bf Z}^2} a_\ell (y)\otimes V^{-a\ell_1} U^{\ell_2} \;
\omega^{-a\ell_1\ell_2 /2}\,
\;e^{-2\pi i \ell \cdot x / b L'},
}
where $\omega \equiv e^{2\pi i/b}$ and
the pair  $U$ and  $V$ are the standard clock and
shift matrices of $SU(b)$ satisfying:
\eqn\tw
{U \; V = \omega\;V \; U.}
The connection $A_{\phi'}$ is a constant-curvature  abelian gauge field
in the diagonal $U(1)$ subgroup
 of $U(N')$ that  satisfies $\phi' = dA_{\phi'}$.
For our particular choice of periodic noncommutative connection
$A$, the effect of $A_{\phi'}$ is to  cancel the first Chern
class induced by $A'$:
\eqn\canc{
\int_{{\bf T}^2} \tr (F' + dA_{\phi'}) =\int_{{\bf T}^2} \tr(F' + \phi') =0.}

 The traceless part of the
ordinary connection $A'$ furnishes a twisted bundle of $SU(N')/{\bf Z}_{N'}$
with 't Hooft magnetic flux $[m'] =r N$ (mod $N'$) and  periodicity  conditions
\eqn\perix
{A' (x_j + L') = \Gamma_j \;A' (x_j) \;\Gamma_j^\dagger.}
The twist matrices may be chosen as
\eqn\tm{
\Gamma_1 = {\bf 1}_N
\otimes U^r, \qquad \Gamma_2 = {\bf 1}_N \otimes V.}

Notice that the {\it same} matrix  Fourier components  $a_\ell$
 appear  in \peri\ and \ord.
 Therefore, when expressed in terms of the $a_\ell$,
 both the action and the integration
measure in the path integral
 remain formally invariant under the Morita transformation, which
acquires the simple interpretation of a change of variables in the
position-space representation. For our example of a periodic
noncommutative two-torus we have the identity
\eqn\larga{
\eqalign{
{1\over 2g^2}
 \int_{\CM_\theta}  \tr \,|F|^2 & + {i\vartheta \over 8\pi^2}
 \int_{\CM_\theta}
\tr \,F \wedge F \cr &= {1\over 2g'^2}
 \int_{\CM'} \tr \,|F'+\phi'|^2 +
{i\vartheta' \over 8\pi^2} \int_{\CM'} \tr \,(F'+\phi') \wedge (F' +
\phi'),
}}
with
$$
\CM_\theta
={\bf T}^2_\theta (L) \times Y, \qquad \CM' ={\bf T}^2 (L') \times Y,
$$
and
the couplings mapping according to the rules:
\eqn\couplm{
g'^{2} \,N' = g^2 \,N, \quad  \vartheta' = \vartheta\,b, \quad \Phi' =
-{r\over b}
.}
Similar manipulations can be performed for other fields in the
adjoint representation of the gauge group.

We assume that this Morita equivalence is a true  isomorphism of
the physical Hilbert space of the theory that, in particular,
 preserves the exact finite-volume spectrum,
including non-perturbative dynamical scales related, for example, to
confinement. In this case, the proper interpretation of \mordu\ is
as a mapping of {\it bare} parameters at some cutoff scale $\Lambda_{\rm UV}$
 that
remains fixed under the duality. A convenient regularization that respects
the duality at the short-distance scales is to consider a $\CN=4$ SYM
theory with appropriate mass terms at the scale $M_s
=\Lambda_{\rm UV}$. In this
case, the map \mordu\ applies to the $\CN=4$ microscopic parameters.

 Thus,
 barring possible
`Morita anomalies', \mordu\  should  hold non-perturbatively  for the continuum
theory.
In $\CN=4$ NCYM theory, the explicit computation of the spectrum
 of $1\over 4$ and $1\over 8$ BPS states by a number of groups \refs{\rbps,
\rmortd}
gives   a Morita-invariant result. In particular, the energies of
 abelian electric and magnetic fluxes are Morita-invariant.
Based on the evidence provided by these examples,
in the rest of this paper we assume that Morita equivalence is free
from anomalies and holds as a true quantum symmetry of the continuum
theory.

\newsec{Rational Approximations and Singular Large-$n$ Limits}

\noindent

Given an infinite convergent
sequence of rational numbers determining  some
 noncommutativity parameter: $\Theta_n \rightarrow \Theta$, we may
define the   noncommutative theory arising as a limit of the Morita-dual
ordinary theories, provided that this limit actually exists. The existence
of this limit is a necessary condition for the
 smoothness of the physics as a function
of $\Theta$. We symbolically denote the resulting limiting theory
by
\eqn\res{
\lim_{n\to \infty} U(N, \Theta_n) \equiv {\overline{U(N)}}_\Theta.
}
Writing $\Theta_n = a_n /b_n$, with $(a_n, b_n) =1$ (i.e. relatively
prime,)\foot{The symbol $(a,b)$ for any two integers $a, b$ represents
their greatest common divisor.} and $b_n >0$,
 we determine appropriate Morita transformations to ordinary
theories by defining numbers $r_n, s_n$ with the property $s_n b_n +
r_n a_n =1$ and performing the transformation \mordu\ for each value
of $n$:
\eqn\otmordu{
T_n = \left(\matrix{ s_n & r_n
 \cr -a_n & b_n \cr}
\right).}
For given values of $a_n$ and $b_n$, we can fix the freedom in the choice of
the pair $r_n, s_n$ by picking $0\leq r_n < b_n$.  

  After Morita duality, we obtain a series of $U(N'_n)$ models with rank
and magnetic flux:
\eqn\rmf{
N'_n = b_n\;N, \qquad m'_n =  r_n \; N,}
where we have assumed that the initial magnetic flux of all
$U(N,\Theta_n)$ models vanishes, a condition that may be enforced
by an appropriate Morita transformation.

Since $b_n \rightarrow \infty$ in the limit, any such rational
approximation is a large-rank limit. However, since the length
of the ordinary tori also shrinks by
\eqn\sh{
L'_n = {L \over b_n},}
we actually have a particular {\it singular} large-$n$ limit, in which
the large-rank limit  is
correlated with the scaling of a two-torus to zero size. One can say
that the
dynamics of the
model ${\overline{U(N)}}_\Theta$   {\it resolves}
  such
singular limits of ordinary gauge theories. In principle, it is not
guaranteed that the model $\ubar$ exists in all situations, in the
sense of all gauge-invariant operators having smooth matrix elements
in the limit. The evidence from BPS sectors in $\CN=4$ NCYM suggests
that such limit makes sense at least for sufficiently supersymmetric theories.

There is evidence  that
$\Theta$-smoothness, if true, must involve rather non-trivial dynamical
phenomena. In  Refs \refs{\rHI, \rirr}, Morita duality was used
as a tool to provide an optimum set of quasilocal descriptions of
the physics of $\CN=4$ NCYM theories on a torus. A `quasilocal description'
is defined by requiring that the elementary degrees of freedom are well-contained
in the box, taking into accout both the quantum size given by the Compton
wave-length and the `classical size' given by the `dipolar extent' of
noncommutative particles: the effective size scales with energy as
$\ell(E)_{\rm eff} = {\rm max}\,(1/E, E\,\theta)$. Demanding $\ell(E)_{\rm eff}
<L$ gives the definition of a given quasilocal patch $1/L < E < 2\pi/\Theta L$.

It is found in \refs{\rirr} that  the structure of quasilocal patches depends
sensitively on $\Theta$. Not only it depends
on the rational or irrational character of $\Theta$, but even
on the `degree of irrationality' according to some well-defined criteria.
Although some non-BPS quantities such as the entropy in the planar approximation
behave smoothly as a function of $\Theta$, it is much less clear that the
generic non-BPS physical quantity (particularly at the level of
non-planar corrections) will behave smoothly given the `multifractal'
nature of the renormalization-group flows.

Ignoring for the moment these caveats, we can entertain
a strong form of the smoothness hypothesis  (see
\refs\rzack) and conjecture that, in appropriate situations,
 the $\ubar$ model is actually equivalent
to some other {\it ab initio} definition of the $U(N)_\Theta$ theory. This
strong form of the conjecture severely constraints the large-$n$ limit
of the ordinary gauge theories. Unfortunately, it concers a singular
large-$n$ limit, and it is unclear what practical information could
be extracted for the {\it regular} large-$n$ limit of 't Hooft \refs\rlargen.

 For
irrational $\Theta$, we can use the $\ubar$ theory as a tentative
non-perturbative definition of the model. In this case, the best rational
approximations are given by  continued fractions. 
For rational $\Theta$, the $U(N)_\Theta$ is itself Morita-dual to some
finite-rank gauge theory. Therefore, in this case, the putative limiting
theory can be rigurously defined, and the equivalence of $\ubar$ and
$U(N)_\Theta$ imposes particularly strong constraints.

Perhaps the most radical statement of $\Theta$-analiticity along these
lines  would be
the one associated to the sequence $\Theta_n = 1/n \rightarrow 0$ in a
rank-one model. In this
case, the series of Morita-dual
 ordinary theories consists of  $U(n)$ models with
one unit of magnetic flux on a torus of volume $L_e L^2 /n^2$. The limit
would be equivalent to an ordinary
 $U(1)$ theory on a torus of volume
$L_e L^2$, i.e. a free theory.
If true, such an equivalence is  rather surprising in view of
the   UV/IR effects discovered in \refs\ruvir, which tend to render
the theory non-analytic around  $\theta =0$.

In the remainder of this paper we evaluate the case for the equivalence
$U(N)_\Theta \equiv \ubar$ from different points of view. Notably, we
use an appropriately defined Witten index as a useful criterion for
$\CN=1$ supersymmetric theories, and the dependence on the instanton
angle as a (less robust) criterion for non-supersymmetric theories.

In the remainder of this section, we collect a number of observations
on the qualitative physics of the $\ubar$ limiting models.
As a final remark, if we try to define the noncommutative gauge theories
using a lattice formulation along the lines proposed in \refs\rnbi,
it seems that one would end up with a prescription similar to the
limiting procedure described in these sections.  According to \refs\rnbi\ 
in their lattice formulation one is forced to have $\Theta$ rational,
thus if we are interested in studying a theory with an irrational value
of $\Theta$, as we take the continuum limit we should also describe a
rational sequence depending on the lattice spacing converging to
the value of interest.  Whether this is the only way to define 
noncommutative theories on the lattice is an open question.

\subsec{Cutoffs and Dynamical Scales}

\noindent

We can define two variants of the singular large-$n$
limits, depending on whether  the
ultraviolet scale $\Lambda_{\rm UV}$ remains fixed in the large-$n$
 limit, or it is scaled appropriately. If $\Lambda_{\rm UV}$ remains
fixed, eventually $L'_n \Lambda_{\rm UV} <1$ and the small torus
shrinks below the cutoff scale. In this case we must interpret
$\Lambda_{\rm UV}$ as a $\CN=4$ supersymmetry breaking scale, $M_s$, so
that the short-distance definition of the ${\overline{U(N)}}_\Theta$
model refers to the $\CN=4$  theory.

Alternatively, we can take the continuum limit $\Lambda_{\rm UV} \rightarrow
\infty$ for each $n$ and define the large-$n$ limit of the series of
continuum field theories. In this case, with $\CN=0,1$ supersymmetry, we
have asymptotically free theories for which the microscopic coupling parameter
transmutes into a dynamical scale $\Lambda_n$. Since the value of each
$\Lambda_n$
is adjustable, we can specify its large-$n$ limit $\Lambda_\infty$ as part
of our specification of the limiting procedure.
One possible uniform definition is given by the localization of
the one-loop infrared Landau pole in the planar approximation:
\eqn\rescale
{
  \Lambda_n = \Lambda_{\rm UV}\,{\rm exp}\left(-{8\pi^2 \over \beta_0
\,
\lambda_n (\Lambda_{\rm UV})} + \cdots\right),}
where the dots stand for higher (planar) loop contributions.
In this formula, $\beta_0 $ is the (positive) one-loop beta function coefficient
with normalization
\eqn\onel{
\beta_0= {11-2n_f - n_s\over 3}
,}
for a theory with $n_f$ Majorana fermions and $n_s$ complex scalars,
all in the adjoint representation of the gauge group,
 and $\lambda_n (\Lambda_{\rm UV}) = {g'_n}^2 (\Lambda_{\rm UV}) N'_n$
is the {\it bare}
 't Hooft coupling of the $U(N'_n)$ theory. Under Morita duality,
the 't Hooft coupling is invariant ${g'}_n^2  N'_n = g^2 N $, with
$g^2 (\Lambda_{\rm UV})$ the bare coupling of the ${\overline{
U(N)}}_\Theta$ theory. In the limit, $\Lambda_n$ converges to
$\Lambda_\infty$,
the standard dynamical scale of the $U(\infty)$ gauge theory arising in
't Hooft's planar limit.

It is important to realize that, despite  the  two-torus
 of vanishing volume ${L_n'}^2
\rightarrow 0$, there is no dimensional reduction on the series of
ordinary theories,
 i.e. the low-energy
 physics
is not `trivialized' into a $(1+1)$-dimensional renormalization-group flow.
The reason is that all of the ordinary gauge theories in the series
have non-zero magnetic flux through the vanishing torus, so that they
support `light' electric-flux
excitations down to  energies  of order
 $1/L'_n b_n = 1/L$, which remains fixed in the limit, and is the
true infrared threshold for the transition to $(1+1)$-dimensional physics.
 The interactions of these
light delocalized modes between the high scale $b_n /L$ and the low scale
$1/L$ are governed by a  renormalization-group
flow with beta function coefficient $\beta_0$ \refs\rafree\ (see also
\refs\rarmoni.) Therefore, asymptotically
free theories with   $L \Lambda_\infty \gg 1$ can develop
  strong coupling before
  the reduction to $(1+1)$-dimensional dynamics takes place.

In summary, the physics of the $\ubar$ models can be very rich, including
possible non-perturbative phenomena at intermediate scales $\Lambda_\infty$.
It must be emphasized that
 our heuristic reasoning is based on the planar one-loop approximation
to the renormalization-group flow of the Yang--Mills coupling. In principle,
non-planar diagrams can yield important contributions
 at energy scales
of order $1/\sqrt{\theta}$. For example, non-planar one-loop diagrams
in the non-compact theory
turn the screening behaviour of the global $U(1)$ modes into antiscreening,
c.f. \refs{\rfernando, \rdurham}. In this case, one shoud further impose
the hierarchy $L\,\Lambda_\infty \gg L/\sqrt{\theta} \gg 1$.

\subsec{Decoupled Photons}

\noindent

Each of the rational $U(N,\Theta_n)$ theories in the
approximating series $\Theta_n \rightarrow \Theta$
  has a free ordinary Maxwell   field
effectively living on a torus of volume $V_n' = V/b_n^2$.
This is a general property of any rational noncommutative theory
on a  finite torus, as  follows from the structure of the vertices in \refpl.
Since momenta are quantized as   $k_i = 2\pi n_i /L_i$
and  $\theta^{ij} = \Theta^{ij} L_i L_j /2\pi$, we
obtain for the symmetric structure constants
\eqn\vanu{
d^{\alpha\beta\gamma}
\; {\rm sin} \left(\pi n_i \Theta^{ij} n'_j\right)
,}
so that  modes in the diagonal $U(1)$ subgroup of $U(N)$ and with momenta
 given by integral multiples  of $b_n$
 decouple.
In the language of the ordinary $U(N'_n)$   Morita dual,
the free $U(1)$ is simply the diagonal subgroup of $U(N'_n)$.
Moreover, since $b_n \rightarrow \infty$,
the excitation gap for these free photons diverges in the
limit $\Theta_n \rightarrow \Theta$, so that they are
irrelevant for the ${\overline{U(N)}}_{\Theta}$ theory at finite volume.
The decoupled photon modes at exceptional momenta $k_i = 2\pi \ell_i b_n /L_i$
are responsible for ultraviolet divergences in non-planar diagrams, unlike
the case of irrational theories in perturbation theory (c.f. \refs\rwulk).
Notice, however, that the infinite gap that is generated in the irrational
limit for these free photons should render the perturbation theory of
${\overline {U(N)}}_\Theta$ equivalent to the standard irrational perturbation
theory.

In the Hamiltonian formalism, the decoupled $U(1)$ contributes an
 energy
\eqn\uuno{
E_{U(1)} = E_\gamma + E_{\rm flux},}
where $E_\gamma$ is the energy in photons and $E_{\rm flux}$ denotes
the contribution of electric and magnetic fluxes, which reads,
 in Morita-covariant
form \refs{\rmortd, \rbps}:
\eqn\fluxes{
E_{\rm flux} = E_e + E_m = {g^2 \over 4N_\Theta} \sum_i {L_i^2 \over V} (
\Theta^{ij} p_j + w^i)^2 +
{1\over g^2 N_\Theta} \sum_i {L_i^2 \over V} \,
(2\pi m^i + 2\pi\,N_\Theta\,\Phi^i)^2 ,}
where $N_\Theta = N+ \half \Theta^{ij} m_{ij}$ is the dimension of the
noncommutative module and $g^2$ stands for the bare coupling  at
the cutoff scale $\Lambda_{\rm UV}$ (the $\CN=4$
high-energy coupling for sofltly broken models.)
 This expression is Morita-invariant and depends
smoothly on $\Theta$. However, for asymptotically free models we take
$g^2 (\Lambda_{\rm UV}) \rightarrow 0$ in the continuum limit, yielding
a divergent magnetic
energy for $m^i + N_\Theta \Phi^i \neq 0$. Thus, in
many of the applications, we shall assume that the $\ubar$ model
has $m^i + N_\Theta \Phi^i =0$. Since we can set $m_{ij} =0$
by an appropriate Morita transformation, we can  assume with no
loss of generality that $\Phi^i= m^i =0$. For the series
of ordinary  $U(N'_n)$ theories in \rmf, we have
\eqn\simp{
m'_n + N'_n \,\Phi'_n =0.}
Notice that, under the same conditions: $g^2 (\Lambda_{\rm UV})
\rightarrow 0$, the contribution to the energy from  abelian electric fluxes
becomes degenerate for all values of the electric flux.

\subsec{Rational Constructions of Theories on ${\bf R}_\theta^3$}

\noindent

A variant of the  limit discussed here can be introduced to provide
 a `purely rational' definition of  a noncommutative
theory on ${\bf R}^3_\theta$ by a blow-up of ${\bf T}^3_\theta$. Consider
a sequence of rational theories $U(N,\Theta_n)$ with $\Theta_n = 1/n
\rightarrow 0$ on ${\bf R} \times {\bf T}_\theta^2 $,  where the
noncommutative tori have size
$L_n = \sqrt{2\pi n \theta} \rightarrow \infty$.
 The commutative description involves a
series of $U(nN)$ ordinary theories defined on shirinking two-tori
of size $L'_n = \sqrt{2\pi \theta /n}$ and supporting exactly  $N$-units
of magnetic flux. In terms of the series of ordinary theories, this
limit is even more singular than the previous one, since the perturbative
gap of electric fluxes vanishes as $\CO(1/\sqrt{n})$.

This `rational' definition of the non-compact noncommutative models may
be useful in discussing the non-perturbative physics associated to
various UV/IR phenomena \refs{\ruvir, \rsut, \rgubser,
 \rgomwise, \rfernando, \rdurham,
\respe}.

\subsec{Large $N$ versus Large $\theta$}

\noindent

It is somewhat dissapointing that the simplest criterion of $\Theta$-analyticity
imposes constraints on singular large-$n$ limits, rather than the ordinary
large-$n$ limit of 't Hooft \refs\rlargen. Thus, an interesting variation of
the limits constructed here would involve stabilizing the size of the
shrinking commutative torus by hand. We can simply consider a combination
of the $\Theta_n \rightarrow \Theta$ limit in the $U(N,\Theta_n)$ models,
together with   a large-volume scaling of the noncommutative torus,
$L_n \rightarrow \infty$. Since $L'_n = L_n /b_n$,
we must define
the sequence of rational  noncommutative theories on tori whose size, $L_n$,
 grows
at least linearly with
  $b_n$.

 More specifically, consider a ${\overline{U(1)}}_\Theta$ limiting
model on growing tori of size $L_n = L' \, |b_n|^\alpha$ with $\alpha >1$, and
$L'$  a fixed length scale.
The Morita-dual tori  have sizes $L'_n = L'\, |b_n|^{\alpha-1}$, which
grow in the  limit. Hence, in this case we end up with exactly
the dynamics of an ordinary $U(\infty)$ theory on
${\bf S}^1_e \times {\bf R}^2$, with a free photon and a tower of
free glueballs at the confining scale  $\Lambda_\infty$.
This rather standard large-$n$ spectrum is encoded in the variables of the
noncommutative theory in a very nonlocal way,
since  the noncommutativity
scale diverges as
$\theta_n \sim  L_n^2 \Theta_n  \sim b_n^{2\alpha} L'^{\,2} \Theta
 \rightarrow \infty$.
 This is a simple way of using Morita duality to argue that
the ordinary large-$n$ limit is captured by a $\theta\rightarrow \infty$
limit \refs{\rtek, \rtoni, \ruvir, \rrabisan}.

Hence, we find that the $\Theta$-analyticity criterion, when forced
to reproduce a standard large-$n$ limit, relates it  to  an extremely
non-local description of physics.  As an illustration of the 
arbitrariness of these limits, let us consider the marginal case
of the scaling above with $\alpha=1$.
The resulting noncommutative geometry is superficially identical to  that
 of $\alpha >1$, i.e.  ${\bf S}^1_e \times {\bf
R}^2_\infty$. However, the Morita-dual tori have fixed volume
$L'_n =L'$, so that  we still have confining phenomenology
on a large box, provided $L'\,\Lambda_\infty \gg 1$,
 except that now we also have a photon with a discrete spectrum
of gap $1/L'$. This is rather exotic when viewed from the point of view
of the limiting  noncommutative theory, since the length scale $L'$ is not
directly associated to the limiting geometry.
Thus, we learn that
what could be called `the $U(1)_\Theta$ theory on
 ${\bf R}^2_\infty$'
is not unique, having at least a one-parameter family of theories labelled
by the hidden scale ratio $L'\Lambda_\infty$.

 In fact, it is likely
that this marginal limit model with $\alpha =1$ is not well defined in the
absence of supersymmetry. Although the
energy scale of the finite box $1/L'$ is largely irrelevant to the
dynamics of the $U(\infty)$  glueballs provided $\Lambda_\infty L' \gg 1$,
the (exponentially small) finite-size effects in this regime
can be  quantified by
 't Hooft's dual  confinement criterion in terms
of screening of magnetic-flux energy. Under very general assumptions,
the non-abelian contribution to the magnetic-flux energy on a large
box is given by (c.f. \refs\rthooft):
\eqn\efl{
E(m_n')_{SU(N'_n)} = C_n \, {L_e \over \Delta^2} \;{\rm exp}\left(
-\sigma\, L'^{\,2}\right)
,}
where $\Delta \sim \Lambda_\infty^{-1}$ is the effective thickness  of the
large-$n$ confining string and $\sigma \sim \Lambda_\infty^2$
 is the corresponding string
tension. The numerical constant is
\eqn\nccc{
C_n = 1-{\rm cos}\,(2\pi m_n' /N_n') = 1-{\rm cos}\,(2\pi r_n /b_n).
}
This quantity is well-defined in the $n\rightarrow \infty$ limit
provided the trigonometric constant $C_n$ has a limit. In general,
this is not the case for arbitrary sequences $a_n /b_n$. One can easily find
examples of approximations to irrational numbers, for which the ratio
$r_n /b_n$ behaves erratically as $n\rightarrow \infty$.  
 In these cases,   there are exponentially small quantities
that are not well-defined in this limit.
This conclusion can be extended to any physical quantity whose
commutative evaluation depends explicitly on the ratio $m'_n /N'_n$ (see
also \refs\rzacku.)

\newsec{The Witten Index}

\noindent

Having established that rational approximations of irrational theories,
denoted ${\overline{U(N)}}_\Theta$,
serve as a `blow-up' of a specific singular large-$n$ limit of ordinary
theories, we would like to study this limit in more detail. In supersymmetric
theories, a significant amount of non-perturbative information is obtained
by looking at the subset of BPS-saturated states. The
most primitive BPS quantity is perhaps
 the supersymmetric (Witten) index, $\Tr (-1)^F$,
 that
counts the number of Bose minus Fermi vacuum states \refs\rwitclass.
 This index gives
valuable  information, not only
 about dynamical supersymmetry breaking \refs\rwitclass, but also
about the dynamics of confinement and the breaking of various global
symmetries \refs{\rwittd, \rwitfd}.
For example, for $\CN=1$ pure Yang--Mills theories in four dimensions
 with gauge
group $SU(N)$ one has $I=\Tr (-1)^F =  N$, reflecting
the spontaneous breaking of the  ${\bf Z}_{2N}$ R-symmetry acting on
the gluinos, $\lambda \rightarrow e^{i \pi  /N} \lambda$, to the
${\bf Z}_2$ subgroup of $2\pi$-rotations: $\lambda\rightarrow -\lambda$.
One can also define refinements of the index that depend on
electric and magnetic fluxes through a torus, and  probe the standard
hypotesis about confinement dynamics, i.e. the confinement of
electric flux, and the screening of magnetic flux \refs\rthooft.

In the large-$N$ limit, the number of confinining vacua is infinite. Naively,
this
 reflects the restoration of the classical $U(1)_R$
symmetry.
Since the classical $U(1)_R$ symmetry was broken to ${\bf Z}_{2N}$ by
instantons,
 the restoration of the continuous symmetry is compatible with the
general --and sometimes naive-- idea that
 instanton effects turn-off in the large-$N$ limit.
Here too the situation is subtle, since domain walls separating adjacent
vacua still have divergent $\CO(N)$ tension (c.f. \refs\rshif).

We would like to test the hypotesis of $\Theta$-analyticity in minimal
supersymmetric theories by computing the Witten index of the limiting model
${\overline {U(N)}}_\Theta$, defined as a limit over the rational
series of $\CN=1$ theories
\eqn\defw
{ I_{\ubar} = \lim_{n\to \infty} I_{U(N, \Theta_n)},
}
where $I_{U(N,\Theta_n)}$ is defined in terms of the index of the
 ordinary $U(N'_n)$
 Morita dual. Incidentally, although this quantity is an integer,
if we do not have continuity in $\Theta$ there is no reason why
the index should not jump in an uncontrollable way as we move
along the sequence $\Theta_n$ approximating $\Theta$. This makes
the computation of the index meaningful and less trivial than
it might naively seem at first sight. 
Before aplying these results to our problem
 we must sort out some technical issues related to the
proper treatment of the diagonal $U(1)$ subgroup of each ordinary
$U(N'_n)$ theory.

\subsec{A Supersymmetric Index for $U(N)$ Theories}

\noindent

In $U(N)$ gauge theories,
 the naive definition of the index as $\Tr (-1)^F$,  in terms of a
trace over the full Hilbert space of the $U(N)$ theory,  gives
a trivial vanishing result from the contribution of the ground
states in the diagonal $U(1)$ sector, obtained by the action
the photino
 zero-mode operators, $\int \tr \lambda$, on the
 vacuum. In fact, the vanishing of the index for a theory with a
$U(1)$ factor is associated to the possibility of adding a Fayet--Iliopoulos
term  that breaks supersymmetry \refs\rwitclass.

The  usual remedy in the study of $\CN=1$  $U(1)$ theories
is to consider instead the refined index $\Tr C (-1)^F$, where $C$ is
the charge-conjugation operator.  Such an index can be defined precisely
when the Fayet--Iliopoulos term is zero, and  is nonvanishing
for the free $U(1)$ theory.  This solution is not directly
 applicable in our case
because, although the charge conjugation is a  symmetry of the $U(N'_n)$
theories, it acts non-trivially on the magnetic fluxes $m'_n
\rightarrow -m'_n$. Thus, the corresponding index is again trivially
vanishing for a generic value of the magnetic flux $m'_n$.

On the other hand,
the physical excitations of the decoupled diagonal $U(1)$  multiplet
are rather uninteresting and even become infinitely massive in the
$n\rightarrow \infty$ limit. Roughly speaking, the Hilbert
space of a $U(N)$ theory
 factorizes between $U(1)$ excitations and $SU(N)$ excitations, the
latter being the interesting ones in our singular large-$n$  limit. Therefore,
it shoud be possible to find a  refinement of the standard  index so that
it does not vanish and gives information about the degeneracy of
ground-states in the interacting (non-abelian) sector.

The details of the factorization of the diagonal $U(1)$ subgroup are
complicated by the global structure of the  gauge group:
 $U(N) = (U(1) \times SU(N))/{\bf Z}_N$. This has two main consequences
for the structure of the canonical quantization on a torus. First,
a given
 $U(N)$ bundle over ${\bf T}^3$, with fixed
  integer-valued first Chern classes $m_{ij}\in {\bf Z}$, can be seen as
a particular $U(1) \times (SU(N)/{\bf Z}_N)$ bundle with quantized magnetic
fluxes $m$ on the $U(1)$ and 't Hooft flux $[m]\equiv m\,({\rm mod}\,N)$ on
the adjoint factor  $SU(N)/{\bf Z}_N$.

A second, related subtlety is that the factorization of gauge transformations
into a diagonal $U(1)$ part and an $SU(N)$ part is ambiguous by elements
of the center ${\bf Z}_N$ of $SU(N)$:
\eqn\fac{
U = e^{i\alpha} \,\cdot\,U' = e^{i\alpha} \,z\cdot z^{-1} \,U',}
where $z^N =1$ and $U' \in SU(N)$. Since ${\bf Z}_N$ is a discrete set,
it follows by continuity that  
this ambiguity only affects the periodicity conditions. In particular,
a strictly periodic $U(N)$ transformation can be factorized into
non-periodic factors satisfying:
\eqn\nonp{
e^{i\alpha(x+L_j)} = z_j \,e^{i\alpha(x)}, \qquad U'(x+L_j) =z_j^{-1}
\; \Gamma_j \,
U' (x) \,\Gamma_j^\dagger,}
where we have considered arbitrary twisted boundary conditions on ${\bf T}^3$.
The entangled action of ${\bf Z}_N$ means that we cannot simply factorize
the Hilbert space into $U(1)$ and $SU(N)$ parts, even at a fixed
value of the magnetic flux.  Rather, we should quantize the theory
{\it without} explicitly dividing by the gauge transformations \nonp, and
then impose the ${\bf Z}_N$-invariance at the end by an averaging procedure.
This is facilitated by the fact that \nonp\ act as global symmetries
on the Hilbert space obtained by satisfying the Gauss law  within the space of
$z_j =1$ gauge transformations.

The characters of the ${\bf Z}_N$ action on the Hilbert space of the
$SU(N)/{\bf Z}_N$ gauge theory define the 't Hooft electric fluxes: $[w]$,
a three-vector of integers modulo $N$. If we construct the
$SU(N)/{\bf Z}_N$  Hilbert space in terms of gauge-invariant Wilson-loop
operators, each component of 't Hooft electric flux corresponds
to the action of a Wilson line, $W'([w^j])$, wrapping the
 $j$-th torus direction and
carrying an irreducible representation with `$N$-ality' equal to $[w^j]$.
We may now add the diagonal $U(1)$  gauge field $
(\tr A)/N$ and build the  general operator:
\eqn\mgen{
W'([w^j]) \, \times \,{\rm exp}\left(i\,w^j\,\oint_j {1\over N}\,\tr A\right)
.}
The integer $w^j$ is interpreted as the abelian electric flux in the $j$-th
direction. Invariance under \nonp\ imposes the expected
constraint between abelian and non-abelian electric fluxes:
\eqn\cons{
[w^j] = w^j\;\;\; {\rm modulo}\;\;\;N.
}

  Hence, we learn that the complete Hilbert space can be constructed
by a decomposition with respect to  value of the total electric flux:
\eqn\fachil
{\CH(w,m)_{U(N)} = \CH (w, m)_{U(1)} \otimes \CH([w], [m])_{SU(N)/{\bf Z}_N},}
which induces  a general decomposition of the $U(N)$ index:
\eqn\gendec{
I(m)_{U(N)} =\sum_{w\in {\bf Z}} I(w,m)_{U(1)} \;\;
I([w], [m])_{SU(N)/{\bf Z}}.}
Thus, the vanishing of the index is a trivial consequence of the vanishing
of all the $U(1)$ factors, even for vanishing electric flux.
In order to include the effects of the instanton angle in more detail,
it  will be instructive to rederive \gendec\ using a path integral argument.

Actually, we can be slightly more general with no extra complications.
Since the covering group of
$U(N)$ is non-compact,  we can define
 a {\it continuous} electric flux valued on a torus, labelling
 representations of the
covering group, ${\bf R} \times SU(N)$,
 that are not representations of $U(N)$. Intuitively, it
corresponds to inserting unquantized $U(1)$ probe charges into the system,
and can be incorporated by adding a topological `theta-term' to the action of
the form
\eqn\eflu{
S_e = {i\over N}\,\sum_j e^j \,\int_{(0j)} \,\tr F_{0j} = {2\pi i\over N} \,
\sum_j e^j \,m_{0j}.
}
With this normalization, $e^j$ are {\it real numbers}
 defined modulo $N$.
The vector of integers $m_{0j} \equiv k_j$ combines with the magnetic
fluxes $m^k = \shalf\,\epsilon^{kij} \,m_{ij}$ to specify the first
Chern classes of the  $U(N)$ bundle over
the euclidean four-torus.  In addition, there is
an integer-valued second Chern class
 that is canonically dual to the standard instanton
angle, entering the action as
\eqn\topac{
S_\vartheta= {i\vartheta \over 8\pi^2} \int \,\tr (F+\phi) \wedge (F+\phi).}
  In this formula, the constant $U(1)$ background field $\phi_{ij}$
 is added for notational
convenience in the applications involving Morita duality. The
extra terms amount to  a constant term
and a redefinition of the electric flux:
\eqn\eee{
e \rightarrow e_\vartheta = e+ {\vartheta\over 2\pi}\;(m+N\Phi).}

The projected  index at a given value of the instanton angle can be
given   a path-integral interpretation
as follows.
\eqn\pint{
I (e,m,\vartheta)_{U(N)} = \Tr \,P_{(e,m,\vartheta)} \,(-1)^F \, e^{-\beta H} =
\CN\,\sum_{k\in {\bf Z}^3} \sum_{\nu\in{\bf Z}}\, e^{-S_e -S_\vartheta}
\;Z(k,m,\nu)_{U(N)},}
where $\CN$ is a normalization constant and
 $Z(k,m,\nu)_{U(N)}$ is the path integral of the $U(N)$ theory (with the
Yang--Mills action)
  on a given
topological sector of ${\bf T}^4$ bundles specified by
magnetic fluxes $m_{ij} =\epsilon_{ijk} m^k$ and  $m_{0i}=k_i$, in addition
to the integral Pontriagin number $\nu\in{\bf Z}$.

On a fixed bundle topology, the partition function
factorizes between $U(1)$ and $SU(N)$ parts:
\eqn\factoo{
Z(k,m,\nu)_{U(N)} = Z(k,m)_{U(1)} \cdot
Z\left([k], [m], \nu\right)_{SU(N)/{\bf Z}_N},}
 where the non-abelian partition function
 depends on $k,m$ only through their mod $N$ reductions $[k], [m]$.
The abelian partition function can be further factorized, according
to \uuno, into the contribution of the free photons (the `Maxwell term')
and that of topological fluxes:
\eqn\faco{
Z(k,m)_{U(1)} = Z_{\rm Maxwell} \cdot e^{-\beta E_m}\cdot
e^{-\pi k\cdot \Omega_e \cdot k}
, \qquad
\Omega_e \equiv {\rm diag}\,\left({4\pi \over g^2 N} {(V_\perp)_i \over \beta
L_i}\right),}
with $(V_\perp)_i$ denoting
 the spatial volume orthogonal to the $i$-th direction, and $E_m$ the
magnetic energy in \fluxes.

The $\vartheta$-dependence of the action $S_\vartheta$ is conveniently
factorized into abelian and non-abelian contributions by the decomposition
of the $U(N)$ curvature:
\eqn\decom{
F= {1\over N}\,\tr\,F + \left(F-{1\over N}\,\tr\,F\right).}
The first term yields a contribution
\eqn\abeco{
{\rm exp}\left[-{i\vartheta \over N}\,k\cdot (m+ N \Phi)\right],}
to the partition function, whereas  the non-abelian contribution is
\eqn\thetadep{
{\rm exp}\left[i\vartheta \left(\nu + {[k] \cdot [m] \over N}\right)\right],}
with $\nu\in {\bf Z}$. The fractional contribution to the instanton charge
in \thetadep\ cancels the analogous term in \abeco, so that the
total second Chern class  $\int
\tr \,(F\wedge F)/8\pi^2$ of the $U(N)$ bundle is  an integer.

Let us separate explicitly the $\vartheta$-dependence induced by
integral instanton numbers by defining the function:
\eqn\nat{
Z([k], [m], \vartheta)_{SU(N)/{\bf Z}_N} =
 \sum_{\nu \in {\bf Z}} e^{i\nu\vartheta} \,
Z([k], [m], \nu)_{SU(N)/{\bf Z}_N}.}
Then, putting all factors together
 we have the following expression for the index:
\eqn\compll{
\eqalign{I(e,m, \vartheta)_{U(N)}
 =& \CN \, Z_{\rm Maxwell}\; \sum_{[k]=1}^{N}
e^{-\pi [k] \cdot \Omega_e \cdot [k]
-{2\pi i \over N}[k]\cdot (e_\vartheta +[m] {\vartheta \over 2\pi} )}\;
Z([k],[m],\vartheta)_{SU(N)/{\bf Z}_N}
\cr &\times \sum_{k' \in {\bf Z}^3} e^{-2\pi i \, k'\cdot e_\vartheta
-2\pi N \Omega_e [k] k'} \times e^{- \pi N^2 k' \cdot \Omega_e \cdot k'},}}
where we have split the sum over $k$ as $k=[k] + N k'$.
We find, upon
Poisson resummation in the integers $k'$:
\eqn\otrcom{
I(e,m, \vartheta)_{U(N)} =  Z_{\rm Maxwell} \sum_{w\in {\bf Z}^3}
\;e^{-\pi ( e_{\vartheta} +  w){1\over N^2 \Omega_e} ( e_\vartheta +  w)}
\;\;\;I
\left( [w]_\vartheta, [m], \vartheta \right)_{SU(N)/{\bf Z}_N} ,}
where
\eqn\omeef{
[w]_\vartheta \equiv
[w]+[m]\,{\vartheta \over 2\pi}.}
In writing down \otrcom,
 we have adjusted the normalization constant so that $\CN^2 = {\rm det}(
\Omega_e)$, and we have introduced the standard index of the $SU(N)/{
\bf Z}_N$ theory
as a function of the 't Hooft electric and magnetic fluxes:
\eqn\deft{
I \left([w], [m], \vartheta \right)_{SU(N)/{\bf Z}_N}
 = {1\over N^3} \sum_{[k]=1}^N e^{2\pi i [k] \cdot
[w]/N} \,Z([k], [m], \vartheta)_{SU(N)/{\bf Z}_N}.}
Notice the appearance of the integrally quantized electric fluxes $w$,
whose reduction modulo
$N$ defines the 't Hooft electric flux of the $SU(N)$ sector.
The effective electric fluxes \eee\ and \omeef\ incorporate the fact,
discovered
in \refs\rwitdyon, that magnetic fluxes induce an electric charge in
the presence of an instanton angle.

The exponential term in \otrcom\ can be recognized as the contribution of
the abelian electric fluxes to the energy,  i.e.
  for an abelian electric flux
of quantum $\varepsilon^i$ we have:
\eqn\eflux{
\beta E(\varepsilon)_{\rm electric} =
 \beta {g^2 \over 4N} \sum_i {L_i^2 \over V}
(\varepsilon^i)^2
= \sum_i {\pi \over N^2 (\Omega_e)_{ii}} (\varepsilon^i)^2.}
 Notice that these energies  become all
degenerate in the continuum limit,
 which involves $g^2 N \rightarrow 0$. However,
the electric fluxes contribute a strictly positive energy in the regularized
theory. We define the index as a limit from the regularized theory, so that
only states with effective abelian flux $\varepsilon=0$ correspond to vacua.

A final compact expression for the  $U(N)$ index  is given as a convolution
of abelian and non-abelian ones that generalizes the formula \gendec:
\eqn\finx{
I(e,m, \vartheta)_{U(N)} =
 \sum_{w\in {\bf Z}^3} I \left( e_\vartheta  + w, m\right)_{U(1)}\;
I \left([w]_\vartheta, [m], \vartheta \right)_{SU(N)/{\bf Z}_N}.}
Thus, the vanishing of the index is due to the trivial vanishing of
the pure $U(1)$ terms, even for zero effective electric flux $e_\vartheta
+ w=0$. This suggests a natural definition of a purely `non-abelian' index
that measures the degeneracy of ground states in the non-abelian sector.
We just formally divide by the purely abelian index at zero effective
electric flux:
\eqn\refin{
I'(e,m, \vartheta)_{U(N)} \equiv {1 \over I(\varepsilon
=0, m)_{U(1)}} \times \;\Tr  \;P_{(e,m,\vartheta)}\; P_{(\varepsilon =0)} \;
(-1)^F\,e^{-\beta H},}
where $\varepsilon = e_\vartheta + w$.
 This formal operation cancels the trivial zero due to the contribution
of photino zero-modes.

In the case that we set $m+ N \Phi =0$ in the initial series of $U(N, \Theta_n)$
models, i.e. we cancel the magnetic vacuum energy, we can give a more
elegant definition of the refined index by twisting the purely abelian
degrees of freedom by the charge-conjugation operator $C_{U(1)}$. We can
specify the action of $C_{U(1)}$ exactly due to the factorization
\fachil:
\eqn\refiin{
I'(e,m, \vartheta)_{U(N)} =
\sfourth \,\Tr \,C_{U(1)} \,P_{(e,m, \vartheta)} \;(-1)^F\;e^{-\beta H}.}
Since the abelian  charge-conjugation operator inverts the sign of
the effective abelian electric flux $\varepsilon \rightarrow -\varepsilon$,
only the $U(1)$ ground states with $\varepsilon=0$ contribute, and we
have a sensible index measuring the degeneracy of non-abelian degrees of
freedom.

We are now ready to use the results of \refs{\rwittd, \rwitfd}. First, notice
that for pure $\CN=1$ $SU(N)$ SYM theories in four dimensions, the classical
$U(1)_R$ symmetry is anomalous, so that we can absorb the $\vartheta$ angle
into a phase redefinition of the gluino field. This means that the physics
will be actually $\vartheta$-independent. On the other hand, the symmetry
$\vartheta \rightarrow \vartheta + 2\pi$ is realized by a permutation of
vacua. The consequence is that, in a sector with both electric and magnetic
fluxes,  electric fluxes differing
 by a redefinition $[w] \rightarrow [w]+[m]$ give equivalent physics, since
this transformation is generated by the $2\pi$-shift in the instanton angle.
This effect, associated to  't Hooft's  mechanism of `oblique confinement'
\refs\robliq,  partitions the $N$ supersymmetric vacua of the $SU(N)$ theory
into $N/c_m$ sets, each one with degeneracy $c_m$, where
\eqn\csubm{
c_m \equiv {\rm max}\;\;([m^i], N), \;\;\;i=1,2,3.}
  Thus,
the result for the $SU(N)$ index is
\eqn\sunin{
I\left([w],
 [m], \vartheta\right)_{SU(N)/{\bf Z}_N} = c_m \;\;\;\;{\rm for} \;\;
[w] = 0 \;\;{\rm modulo} \;\;[m]\;{\bf Z},}
and it vanishes otherwise.

The abelian index $I'(e_\vartheta+ w, m)_{U(1)}$
 vanishes unless the effective electric-flux
energy is zero, which requires $ e_\vartheta =  -w$ or, equivalently,
$e_\vartheta$ must be an {\it integer}.
Combining the two selection rules with the convolution \finx\ we find the
result for our  refined $U(N)$ index:
\eqn\totrul{
I' (e, m, \vartheta)_{U(N)} = c_m \;\;\;{\rm for} \;\;
 e_\vartheta  = 0 \;\;{\rm modulo} \;\;[m]\;{\bf Z},}
and it vanishes otherwise.

In particular, \totrul\ implies that the non-abelian index for a $U(N)$ theory
with periodic boundary conditions $m=0$
 is concentrated at zero electric flux $e=0$ and has value $N$. Thus,
it diverges in the large-$N$ limit.

\subsec{The Supersymmetric Index of the  ${\overline{U(N)}}_\Theta$ Model}

\noindent

We are ready to apply these results to our problem of determining the number
of non-trival supersymmetric vacua of the ${\overline{U(N)}}_\Theta$ limiting
theory. Since the ${\overline{U(N)}}_\Theta$ model was defined with periodic
boundary conditions, a similar behaviour to that of the commutative counterpart
would suggest an infinite index at zero electric flux,
 given the fact that we define the model via
a large-$n$ limit.
 However, this naive
 expectation
is not borne out by the explicit
calculation.
 In our limit definition:
\eqn\deflim{
I' (e)_{{\overline {U(N)}}_\Theta} = \lim_{n\to \infty}
\,I'(e, m'_n)_{U(N'_n)} }
we use the electric flux $e$
 of the ordinary theories $U(N'_n)$ to label the states.
Note that, for the components parallel to the noncommutative directions,
 this is Morita dual to fixing a projector over some combination
of momenta and electric fluxes in the series of noncommutative theories. In
particular, a fixed value of the integer electric flux in the ordinary
$U(N'_n)$ theories maps under the duality to the combination:
\eqn\mapc{
w'_n = b_n\,(w+ \Theta_n \,*p ),}
where $p$ and $w$ are momenta and electric flux on the noncommutative
${\bf T}^2_{\theta_n}$.    Since
we find more transparent
 the language of the ordinary unitary theories, we choose
to parametrize the index in
 terms of the $U(N'_n)$ electric flux $e$. Notice also
that, since $N'_n \rightarrow \infty$, this flux is asymptotically defined
as an arbitrary vector in ${\bf R}^3$ in the large-$n$ limit.

The important point is that $(m'_n, N'_n) = N$ is bounded and constant in the
limit. Hence, the result \totrul\ translates into
\eqn\indref{
I' (e)_{{\overline {U(N)}}_\Theta} = N \;\;\;\; {\rm for} \;\;\;\; e=0 \;\;\;
{\rm modulo}\;\;\; (0,0,N)\;{\bf Z}.}
 Notice that the definition
of the index with the extra phase depending on the rank of the gauge
group, as in \refs\rwitfd,
 would yield an extra factor of $(-1)^{N'_n -1}$, rendering the index erratic
in the large-$n$ limit.

Thus,
the  refinement of the index by the electric flux {\it is} smooth.  The index
 has been `fractionalized' due to the physics of oblique confinement
in the series of ordinary theories. 

The refinement by the electric flux is
essential in obtaining a smooth answer. Since the index is computed in a
Born--Oppenheimer approximation, the result is determined by  the structure of
the space of classical  ground states, which in turn depends   
  sensitively on the
{\it rational} value of $\Theta$.   
 For a periodic  $U(N)_\Theta$ model with   $\Theta = a/b$,
the corresponding moduli space has $b$ connected components, each one
 of dimension
$N$. This is clear in the Morita-dual picture of the gauge bundle,
where we have (for our particular choice of background field), a commutative
$U(1) \times SU(N')/{\bf Z}_{N'}$ bundle with $N' = N b$ and $N r$ (mod $N'$)
 units
of 't Hooft  flux  through ${\bf T}^2$. The flat connections are characterized
by holonomies in the non-abelian factor
 $U_1, U_2, U_e$ satisfying
\eqn\triple{
U_1 \,U_2 = U_2 \,U_1 \;e^{2\pi i r/b}, }
and
\eqn\otrr{
U_e \,U_j = U_j \,U_e , \;\;\;\;{\rm for}\;\;j=1,2.}
The solution has $b$ components
\eqn
\slo{
U_j = (H_j \otimes {\bf 1}_b )\cdot  \Gamma_j, \qquad U_e = H_e
\otimes e^{2\pi i q/b} {\bf 1}_b,}
where $H_j, H_e$ are in the Cartan torus of $SU(N)$ and $q=0, 1, \dots, b-1$.
The different components correspond to the different sectors of 't Hooft's
electric flux that can be related by tunneling via fractional instantons. This
explains why the index is only smooth in the large-$b$ limit once it is
refined at a fixed value of the electric flux. The relation between
the  fractional instantons and the structure of the space of 
 flat connections suggests  
that the contribution of fractional instantons, if non-vanishing,
tends to  work {\it against} continuity in $\Theta$. In the next section we
present a quantitative test of this idea.

\newsec{Instanton-induced  $\vartheta$-dependence}

\noindent

Our treatment of the index in $U(N)$ theories can be formally
generalized to
other physical quantities. For example, if we consider the antiperiodic
spin structure for the gauginos, we are computing the finite-temperature
partition function. This partition function admits a similar factorization
on electric-flux labels:
\eqn\part
{e^{-\beta F(e,m, \vartheta)_{U(N)}} = \sum_{w\in {\bf Z}^3} e^{-\beta F(e+
 w)_{U(1)}} \; e^{-\beta F([w]_\vartheta, [m], \vartheta)_{SU(N)}},}
where now we have a non-trivial  $\vartheta$ dependence
 due to the breakdown of supersymmetry.

 It is often the case that,
particularly when the  instanton-gas approximation suffers from
infrared problems,     the $\vartheta$-dependence
of physical quantities is apparently invariant under $\vartheta \rightarrow
\vartheta + 2\pi N$ only. This is especially obvious when considering a
soft breaking of $\CN=1$ supersymmetry by the addition of a small gluino mass.
This has the effect of splitting the vacuum energies of the $N$ vacua, that
are not equivalent any more. Since $\vartheta \rightarrow \vartheta + 2\pi$
shifts the vacua one by one, we need $N$ steps to return to the same vacuum.

For $SU(N)/{\bf Z}_N$ theories in finite   volume, the fractional nature
of the instanton number means that, in principle, physics is only periodic
in $
\vartheta$ modulo $2\pi N$. However, for the case of   electric-flux energies,
 the $2\pi$ periodicity
of the instanton angle is restored by a non-trivial level-crossing
\refs\rrevbaal.
 The fractional
$\vartheta$-dependence
implicit in the redefinition of $SU(N)/{\bf Z}_N$ electric
fluxes:
\eqn\refd
{[w]_\vartheta = [w] + [m] \,{\vartheta \over 2\pi},}
gives the appropriate spectral flow $[w] \rightarrow [w] +[m]$.

Similar behaviour should be found in the $U(N')$ theories that appear
as Morita duals of rational noncommutative theories, provided the
abelian contribution to $\vartheta'$-dependence cancels out. The condition
for this is the same as the vanishing of the magnetic ground-state energy:
$m'+ N' \Phi' =0$.

Under Morita duality, the instanton angle is fractionalized as
\eqn\fara{
\vartheta' = b \,\vartheta.}
Thus, $2\pi$-periodicity in the $U(N,\Theta_n)$ model translates into
naive $2\pi b_n$ periodicity in the $U(N'_n)$ model. If the physics
is $\vartheta$-dependent, such as in the absence of supersymmetry,
we may have some level-crossing phenomena that recover the $2\pi$-periodicity
in $\vartheta'_n$. As a result, the $U(N,\Theta_n)$
 model would actually show a hidden $2\pi/b_n$ periodicity in the spectrum as a
function of $\vartheta$. Thus, level-crossing phenomena of order $b_n$ could be
 present in rational theories with $\Theta_n = a_n /b_n$.
In the large-$n$ limit we have $b_n \rightarrow \infty$, so that the
$\vartheta$-dependence of the ${\overline{U(N)}}_\Theta$ model would be either
trivial or infinitely discontinuous. In both cases, we would violate
the strongest form of the $\Theta$-smoothness conjecture, since the
$\vartheta$-dependence for rational $\Theta$ must be non-trivial  and
smooth.

 We can test this scenario with an explicit example.
  Let us consider the non-perturbative splittings of
 the electric-flux
energies in the non-supersymmetric theory,
 calculated using a dilute gas of fractional instantons
(see  \refs{\rrevbaal, \rrevtoni, \rusmat}).

We consider the simplest case of a rank-one
rational noncommutative theory   on a periodic torus ${\bf S}^1 (L_e)
\times {\bf T}^2_\theta (L)$, with $\Theta = a/b$.
The ordinary dual $U(N')$ theory lives
in ${\bf S}^1 (L_e) \times {\bf T}^2 (L')$, with $N'=b$, $L' = L/b$.
 We have $m' + N' \Phi' =0$ so
that the diagonal $U(1)$ has effectively zero magnetic flux.
The $SU(N')/{\bf Z}_{N'}$ bundle has 't Hooft flux
$$
[m']= (0,0, [m'_e])= (0,0,r),
$$
where $sb+ra=1$ and $0\leq r <b$.
Then, the non-abelian contribution to the energies of electric fluxes
of the form $[w]=(0,0,[w_e])$ is exactly degenerate to all orders in
perturbation theory. The leading effect lifting this degeneracy is
the tunneling contribution by fractional instantons of charge proportional
to $1/N'$. We can estimate this splitting by a standard computation
in the   dilute-gas approximation.

Let us consider the sum over all   chains
 of $k_+$ fractional instantons and $k_-$
anti-instantons of minimal charge
 along the euclidean time direction,  in the limit $\beta/L'
\rightarrow \infty$. Using $ar+bs=1$, we can write the total instanton
charge in each term in the sum as
\eqn\totch{
Q= {k_+ - k_- \over N'} = {(k_+ - k_-)\, a\, [m'_e] \over N'} + (k_+ - k_-)\, s.}
Using the standard parametrization of the topological charge
\eqn\stpar{
Q= {[k']\cdot [m'] \over N'} + \nu,}
with $\nu\in {\bf Z}$,
we can identify $[(k_+ - k_-)\,a] = [k'_e]$ as the conjugate variable to
the electric flux $[w'_e]$. In particular,
in computing  the discrete Fourier transform
\eqn\disf{
e^{-\beta E([w'], [m'], \vartheta')} = \sum_{[k']=1}^{N'} e^{2\pi i [k']\cdot
 [m'] /
N'} \,\sum_{\nu \in {\bf Z}} e^{i\vartheta' Q}\,Z\left([k'], [m'], \nu\right)
}
within the instanton-gas approximation,
 we can replace the sum over $[k'_e]$ by a free
sum over integers $(k_+ -k_-)$. Assuming the usual factorization of
the instanton measure in the dilute instanton-gas approximation, the sum
over $k_+$ and $k_-$ exponetiates and we find for the instanton-induced
splitting:
\eqn\findd{
\Delta E \left([w'_e], [m'_e], \vartheta' \right) =
 -2\,K\;{\rm cos}\,\left[{1 \over N'} (
2\pi \,a\, [w'_e] + \vartheta')\right].}
Going back to the variables of the original noncommutative $U(1)$ theory,
we notice that $[w'_e]$ is Morita invariant in our particular case, and
we have
\eqn\findc{
\Delta E \left([w_e], \vartheta \right)
 = -2\,K\;{\rm cos}\,(2\pi\, [w_e]\,\Theta +
\vartheta).}
The numerical prefactor $K$ is interesting. It takes the form:
\eqn\omegat{
\beta\,K\propto  \beta \,L_e \, L'^{\,2}\cdot N'^{\,2} \cdot (\Lambda_{\rm
UV})^4 \cdot \left(\sqrt{S_{cl}}\right)^4 \cdot e^{-S_{cl}}
\,\cdot\; \left(\left|{{\rm Det}'_F \over {\rm Det}'_B}\right| +
\dots\right).}
The first  term in this expression
 gives the contribution of the four translational
zero modes of the fractional instanton. There is a degeneracy factor
$N'^{\,2}$ coming from the various independent tunnelings for a given
topological charge.\foot{There are $N'^{\,2}$ inequivalent
instanton solutions on ${\bf R} \times {\bf T}^3$
 that tunnel between fixed pairs of vacua on ${\bf T}^3$. They are
obtained by the action of discrete translations in the plane of twisted
boundary conditions (c.f. \refs{\rmargas, \rrevtoni, \rusmat}).}
  The  terms containing the
classical action
$$
S_{cl} = {8\pi^2 \over g'^2 (\Lambda_{\rm UV})N'}
$$
 include
the   usual
Jacobian  $\sqrt{S_{cl}}$ for each collective coordinate.   A factor
 $(\Lambda_{\rm UV})^4$ comes from the zero-modes in the  regularization
of the one-loop determinants, whose non-zero-mode contribution  for bosons
and fermions is
also indicated. The dots in \omegat\ stand for higher-loop
 contributions.

Let us consider $\Lambda_{\rm UV} = M_s$ as the scale of soft breaking
of $\CN=4$ supersymmetry down to $\CN=0$. In terms of this fixed scale, we
 can rewrite the numerical prefactor in $U(1)_\Theta$ variables as
\eqn\rewri{
K \propto L_e \,L^2 \cdot (M_s)^4 \cdot \left({8\pi^2 \over g^2 (M_s)}\right)^2
\,\cdot\, e^{-8\pi^2 /g^2 (M_s)} \;\cdot \; \left(\left|{{\rm Det}'_F \over
{\rm Det}'_B}\right| + \dots \right).}
Notice that the factor of $(L' N' )^2$ in \omegat\ combines into a single
factor of $L^2$, as corresponds to a single position collective coordinate
for the noncommutative $U(1)$ instanton (c.f. \refs\rncinst)
 that appears as a Morita dual of
the ordinary fractional instanton.

The important property of \findc\ and  \rewri\ is that the size of the
instanton-induced $\vartheta$-dependence is non-perturbative in the
$\CN=4$ coupling $g^2 (M_s)$. This is to be compared to the splittings
induced by the energies of the abelian fluxes \fluxes:
\eqn\abspli{
\Delta\,E(w_e)_{\rm abelian} = {g^2 (M_s) \over 4} \,{L_e \over L^2} \,(w_e)^2.}
Clearly, within the conditions of applicability of the instanton expansion,
the non-abelian contribution is  completely negligible in comparison
to the abelian one. As a result, there are no possible level-crossing
phenomena induced by \findc, and energy levels are clearly continuous
under rational approximations of any given $\Theta$.

The qualitative picture changes considerably if we decouple completely
the $\CN=4$ regularization. Namely, let us remove completely the
$\Lambda_{\rm UV}$ cutoff scale by proper renormalization in the
$g^2 (\Lambda_{\rm UV}) \rightarrow 0$ limit. In this limit, the
abelian splittings \abspli\ vanish, and we are left with the non-abelian
contributions \findc. Therefore, level-crossing becomes possible.

In order to correctly renormalize \omegat\ we must identify the effective
infrared cutoff of the one-loop determinants. This is given by the
size of the noncommutative $U(1)$ instanton  in the noncommutative
box of  volume $L_e L^2$. In the limit that $\sqrt{\theta} \ll L_e, L$,
this size is of $\CO(\sqrt{\theta})$, since the instantons must be
a smooth deformation of the instantons at $\theta=0$. In this case,
the instanton at $\theta =0$ is point-like, and the only available scale
for the resolution is given by  $\theta$.

Therefore, provided the theory is still perturbative at the energy scale
$M_\theta = 1/\sqrt{\theta}$, i.e. $g^2 (M_\theta) \ll 1$, we can eliminate
$\Lambda_{\rm UV}$ in favour of $\theta$ and the dynamical scale
$$
\Lambda = \Lambda_{\rm UV} \;e^{-8\pi^2 /{\beta_0} g^2(\Lambda_{\rm UV})},
$$
with $\beta_0$ given in \onel.  We obtain
\eqn\renor{
K  = L_e\,L^2 \cdot (M_\theta)^4\,\cdot \left({
\Lambda \over M_\theta}\right)^{\beta_0} \;\cdot \;f\left(g^2 (M_\theta)\right),}
where we have summarized in the function $f(g^2)$ the
 renormalized perturbative expansion in the
instanton background.

Consider now a rational approximation of some generic noncommutativity
parameter: $a_n/b_n = \Theta_n  \rightarrow \Theta$. The induced splittings of
electric fluxes are given by
\eqn\indf{
\Delta E\left([w_e], \vartheta\right)_n = -2\,K_n\,{\rm cos}\,
\left(2\pi \,[w_e] \,\Theta_n +
\vartheta\right).}
In the large-$n$ limit, $K_n \rightarrow K_\infty$ with
$$
K_\infty = L_e\, L^2 \,(M_\theta)^4 \,\left({\Lambda_\infty \over M_\theta}
\right)^{\beta_0} \;f\left(g^2(M_\theta)\right), 
$$
and each individual curve for fixed $[w_e]$ converges to the limiting
curve
$$
-2\,K_\infty\,{\rm cos}\,\left(2\pi \,[w_e] \,\Theta + \vartheta\right).
$$
However, the different curves obtained by varying $[w_e]$  cross one another
as a function of $\vartheta$. Therefore, the ground-state  energy as a
function of $\vartheta$ is defined by the minimum:
\eqn\gsi{
E(\vartheta)_{\rm gs} = \lim_{n\to \infty}\;\;\min_{0\leq [w_e] <b_n}
 \;\left[-2\,K_n
\,{\rm cos}
\,\left(2\pi \,[w_e]\,\Theta_n + \vartheta\right)\right].}
If $\Theta$ is irrational, the minima of the cosine function are uniformly
distributed in the limit. Therefore, in this case, the ground-state energy
is a constant, independent of $\vartheta$. In general, let us write
$\Theta_n = \Theta + \Delta\Theta_n$ and $[w_e] = x\, b_n$. In the limit,
$x$ becomes a continuous variable in the unit interval and we have,
\eqn\limm{
E(\vartheta)_{\rm gs} \rightarrow  -2\,K_\infty\; \min_{0\leq x \leq 1}\;
 {\rm cos}\,\left(2\pi \,[w_e]
\,\Theta + \vartheta + 2\pi\, x \,b_n \,\Delta \Theta_n\right).}
For rational $\Theta = a/b$, the direct computation of the resulting
ground-state  energy yields
\eqn\direcc{
E(\vartheta)_{\rm gs}  \propto \min_{0\leq [w_e] <b}\;\;{\rm cos}\,\left(2\pi
\, [w_e]
\,{a\over b} + \vartheta\right).}
Hence,
the condition for \limm\ to approach \direcc\ is that the combination
$$
b_n \,\Delta\Theta_n = b_n \,\left({a_n \over b_n} -\Theta\right) \rightarrow 0
$$
in the large-$n$ limit. However, this is true if and only if  the limiting
value $\Theta$ is an {\it irrational} number. We conclude that, for rational
$\Theta$, the term $2\pi x \,b_n \Delta \Theta_n$ shifts the argument of
the cosine function in \limm\ by an arbitrary amount in the large-$n$ limit,
so that the minimum  is given by $E(\vartheta)_{\rm gs} = -2K_\infty$ for
any $\vartheta$.

Approximations by rationals give a constant ground-state
energy, in disagreement with the direct evaluation \direcc.
This implies that the strongest possible conjecture of $\Theta$-analiticity
over the rationals fails this test. It is very interesting that 
the $\Theta$-discontinuity that we have discussed 
involves a complete decoupling of the supersymmetric energy scale, i.e.
$M_s \rightarrow \infty$ faster than the large-$n$ limit.

\newsec{Models in  $2+1$ Dimensions}

\noindent

Many of the previous results can be extended to the case of rational
approximations of minimal supersymmetric models in $2+1$ dimensions. In this
case we borrow the results of Ref. \refs\rwittd\ for the case of
$SU(N)$ gauge group.
  The  index
is a function of the topological Chern--Simons coupling
 $k$,
 governing the  supersymmetric Chern--Simons mass term
\eqn\cs{
CS(A)_{SU(N)} =
 {k\over 4\pi} \int \tr\,\left(A\wedge dA +
{2\over 3}\; A \wedge A \wedge A + {\bar \lambda} \lambda\right).}
 This quantity
satisfies a topological quantization condition ${\bar k} \in {\bf Z}$ where
${\bar k} \equiv k-N/2$, the shift by $N/2$ being a consequence of the
contribution of the fermionic measure.  The result of \refs\rwittd\
 for
$SU(N)_k$ with $k>0$  and periodic boundary conditions is
\eqn\tresdi{
I(k,N)_{SU(N)} = {(N+{\bar k} -1)! \over {\bar k}!\, (N-1)!}
.}
We also have $I(-k,N) = (-1)^{N-1} I(k,N)$ and
 the index vanishes for $|k| < N/2$.

For the group $SU(N)/{\bf Z}_N$, i.e. allowing 't Hooft's magnetic fluxes
 $[m]\in {\bf Z}_N$
 the quantization condition is instead
\eqn\instead{
k = {N \over 2}
 \;\;\;{\rm modulo}\;\;\;{N  \over (N,[m])} \,{\bf Z},}
due to the fractionalization of the instanton number. 
 In particular,
for such non-trivial gauge bundles the Chern--Simons action must be
defined in terms of a four-dimensional topological action.  Extending
the bundle to an appropriate
 four-manifold $X$ whose boundary is the desired three-manifold
$\CM_3 = \partial X$, we define the Chern--Simons term as:
\eqn\cherns{
 CS(A)_{\CM_3} = {k\over 4\pi} \int_X {\rm tr}\,
\left(F\wedge F
+ d({\bar \lambda} \lambda)\right).}
For example, if $\CM_3 = {\bf S}^1 \times {\bf T}^2$, we can take $X=D\times
{\bf T}^2$, where $D$ is a two-dimensional disk over which the bundle
extends trivially. The fermionic term in \cherns\ does not need
to be extended to the interior of $X$ because the action is already
well-defined on its boundary.

\subsec{Morita Duality}

\noindent

  The definition \cherns\  is suitable
to the study of  Morita
duality in the noncommutative case, since the Morita transformation
of the right hand side is the same as that of the Yang--Mills action
 (after properly redefining the field strength $F\rightarrow F+\phi$).
 Working on $ \CM_\theta = {\bf S}^1_\beta \times {\bf T}_\theta^2 (L)$,
 with periodic
boundary conditions on the noncommutative torus,
 under the Morita duality \mordu\ we have:
\eqn\samec{\eqalign{
{4\pi \over k}\;NCCS(A)_{\CM_\theta}
&= \int_{\CM_\theta} \tr\,\left(A\wedge dA + {2\over 3}
\;A\wedge_\star A \wedge_\star A + {\bar \lambda}
 \lambda\right) \cr  &=
 (s+r\,\Theta) \; \int_{X'} \tr\;\left[(F' + \phi') \wedge (F'+\phi')
+ d({\bar \lambda}' \,\lambda')\right]
,}}
where $X' = D \times {\bf T}^2 (L')$ and the notation $A\wedge_\star A $ means
$A_\mu \star A_\nu \;dx^\mu \wedge dx^\nu$. In the particular case
of interest to us, we have $s+r\,\Theta = 1/b$ and $L' = L/b$.
 Thus, we learn that the T-duality mapping of the
topological Chern--Simons coupling is
\eqn\maptk{
 k\rightarrow k' = k\, (s+r\Theta)^{-1} = k\,b.}
 Notice that the rescaling by a factor of $b$
agrees with the quantization condition in the ordinary $U(N')$ theory, in
the presence of 't Hooft fluxes,
since  the Chern--Simons coupling  of the (periodic) noncommutative theory,
$k$,
is quantized modulo integers \refs\rquank.
 Our normalization
conventions for  the Chern--Simons action, together with the general
action of Morita duality for rational $\Theta$, gives the general
Morita-covariant quantization condition in sectors with arbitrary
magnetic flux $m$,
\eqn\genqua{
k = {N_\Theta \over 2}
 \;\;\;{\rm modulo}\;\;\;{N_\Theta  \over (N,m)} \,{\bf Z},}
where $N_\Theta = N-\Theta m$ is the dimension of the noncommutative module.
 This quantization condition
should extend to general irrational $\Theta$ by analytic continuation.\foot{
The quantization condition \genqua\ should agree with that in
Ref. \refs\rkraj, given the proper notational conventions.} This is a
reasonable expectation since all we use to obtain \genqua\ are 
classical properties of the classical action which is analytic in
$\Theta$.

In particular, for our rational
sequence of T-dualities between $U(N,\Theta_n)$ and $U(N'_n)$, we have
\eqn\sectres{
k'_n = b_n \,k,
}
where $k$ is the fixed
 topological coupling in the noncommutative $U(N,\Theta_n)$
theories, which acquires the quantization condition:
\eqn\quanss{
k = {N\over 2} \;\;\;{\rm modulo}\;\;\; {\bf Z}.}
It is interesting to notice that the shift by $1/2$ in the rank-one
case can be understood directly by integrating-out the noncommutative
Majorana fermions \refs\rchucs.

\subsec{Level Normalization and the Born--Oppenheimer Approximation}

\noindent

In the microscopic evaluation of the index one considers the
effective dynamics of the gauge and fermion zero-modes \refs{\rwitclass,
\rwittd, \rwitfd}. Unlike the
four-dimensional case, where the index was independent of the value
of couplings in the Lagrangian,  in $2+1$ dimensions the
index depends on the Chern--Simons level. Therefore,
 its absolute normalization
is essential in the computation.

A naive zero-mode reduction of the noncommutative Chern--Simons Lagrangian
on the noncommutative torus with periodic boundary conditions yields
the effective Born--Oppenheimer Lagrangian (we only consider the bosons
in this discussion)
\eqn\naiv{
{k \over 4\pi} \int_{{\bf R} \times {\bf T}^2_\theta} \tr \;\left( A \wedge
dA + {2\over 3}\; A \wedge_\star A \wedge_\star A\right)
 \longrightarrow {k \over 4\pi} \int dt \;\epsilon^{ij}\;C_i^\alpha \,
\partial_t \,C_j^\alpha
,}
where
\eqn\defc{
C_j^\alpha \equiv \tr \left(H^\alpha\,(a_0)_j \,L\right),}
is defined in terms of
  the constant component $a_0$ of the noncommutative gauge field
in the Fourier expansion \peri. The $N$ matrices $H^\alpha$ are a convenient
basis of the Cartan subalgebra of $U(N)$, normalized as $\tr (H^\alpha H^\beta) =
\delta^{\alpha\beta}$.

After the Morita transformation, in terms of the twisted ordinary
$U(N')$ gauge theory,
 the  flat connections are parametrized simply by those in the  non-abelian
sector\foot{This is a consequence of
 our choice of background field and magnetic
fluxes in the original model,  making the diagonal $U(1)$ effectively
flat \canc.}. The moduli space of
 flat connections of the twisted $SU(N')/{
\bf Z}_N'$ bundle on a two-torus  is isomorphic to that of $SU(N)$ flat
connections on the periodic torus. It can be constructed explicitly
in terms of the embedding $SU(N) \otimes SU(b) \subset SU(bN) = SU(N')$
that is  realized in formula \ord.

  The Born--Oppenheimer reduction to zero modes reads
\eqn\hu{\eqalign{
 {ik' \over 4\pi}
\int_{D\times {\bf T}^2 (L') } \tr\,(F'+\phi') \wedge &( F' + \phi')
\cr &\longrightarrow {ik' \over 4\pi} \int dt \,(L')^2\,\epsilon^{ij}
\,(a_0)_i^\alpha \,\partial_t \,(a_0)_j^\beta \,\tr\,
 (H^\alpha  \otimes {\bf 1}_b)(H^\beta \otimes
{\bf 1}_b).
}}
Using now
$$
\tr \,(H^\alpha \otimes {\bf 1}_b)(H^\beta
 \otimes {\bf 1}_b) = b\; \delta^{\alpha
\beta}
,$$
together with $L' = L/b$ and $k' =  k b $ we find exactly the same
 effective action
\eqn\eflan{
{k \over 4\pi} \int dt \,\epsilon^{ij}\,C^\alpha_i \,\partial_t
 \,C^\alpha_j
,}
when expressed in terms of the angular variables
$C_j^\alpha = (a_0)_j^\alpha \,L$, just as before.
Thus, we learn that, despite the rescaling of the
Chern--Simons level under Morita duality $k\rightarrow k b$, which is
required by our trace normalizations,  the
space of flat connections is sensitive to the same effective Chern--Simons
level before and after the duality. The reason for this is of course
that the Fourier components $a_\ell$ remain the same in both
representations and, in particular, their periodicity should not change.
This is also clear from the fact that the zero modes know little about
noncommutativity.

The periodicity of the zero-mode variables $C_j^\alpha$ is enforced by
 gauge transformations that are periodic on the noncommutative torus
and  satisfy the condition of
Moyal-unitarity: $U\star U^\dagger = U^\dagger \star U = {\bf 1}_N$.
Expanding in Fourier series:
\eqn\expan{
U(x) = \sum_{\ell \in {\bf Z}^2} u_\ell \;e^{-2\pi i \ell \cdot x /L},}
where the coefficent matrices $u_\ell$ are not unitary in general.
It is then easy to check that the twisted gauge transformations
\eqn\twi{
U'(x) = \sum_{\ell \in {\bf Z}^2} \;u_\ell \otimes V^{-a\ell_1} \,U^{\ell_2}\;
\omega^{-a\ell_1 \ell_2 /2} \;e^{-2\pi i \ell\cdot x/L},}
satisfying
\eqn\perrr{
U'(x+L_j) = \Gamma_j \;U'(x) \;\Gamma^\dagger_j,}
enforce exactly the same periodicity conditions
on $C_j^\alpha$ when acting on $A' + A_{\phi'}$, as defined  by
 \ord.

\subsec{The $2+1$ Dimensional Index}

\noindent

The upshot of the discussion in the
 preceding subsection is that, when calculating the
index of $(2+1)$-dimensional gauge theories by quantization of the
moduli space of flat connections, the effective Chern--Simons
level is exactly given by that  of the noncommutative theory, which
remains fixed. Thus, the index is $\Theta$-independent
and can be calculated as if the ${\overline{U(N)}}_\Theta$ model was
an ordinary gauge theory with periodic boundary conditions on the
torus.

This means that, whatever its value, the index will be
only a function of $k$ and $N$, and it is obviously smooth under
rational approximations of $\Theta$. Although this settles our
main concern, it is nevertheless interesting to pursue this matter
and compute the index of the $\ubar$ model.

Following our general discussion \gendec, the
 result factorizes into abelian and non-abelian components for each
value of the electric flux. The linking between the $U(1)$ and $SU(N)$
sectors is done by the  ${\bf Z}_N$  quotient acting on Wilson loops
in  each direction, enforcing the  constraint $[w^i] = w^i$ (mod $N$), 
$i=1,2$.

Since the infrared behaviour of the $2+1$ models under consideration
is dominated by the Chern--Simons term, the degeneracy of the ground
states of the full theory is naturally related to the dimension of
the Hilbert space of the topological Chern--Simons model describing the
infrared limit.

 One subtlety of the description in terms of the effective
topological theory is that 
 Wilson lines wrapping homologically
inequivalent cycles of the torus are canonical conjugates of one another.
Thus, the states are labelled by the eigenvalues of only one set of Wilson
lines.
In the abelian case, this can be understood in elementary terms.
For an abelian Maxwell--Chern--Simons model with action
\eqn\mcs{
S_{U(1)} = {1\over 4e^2} \int |d{\CA}|^2 +{i\kappa \over 4\pi}
\int {\CA}\wedge d{\CA} +
{\rm Fermi \;\;terms,}}
we can explicitly find the zero energy states on the torus with periodic
boundary conditions (see, for example \refs\rpoly\ and references therein).
 Only the holonomies of the gauge field ${\CC}_i =
\oint_i {\CA}$ are important, with effective Lagrangian
\eqn\efla{
\CL_{\rm eff} = {1\over 2e^2} \,(\partial_t \, {\CC}_i)^2 + {\kappa \over 4\pi}
\,\epsilon^{ij} \,{\CC}_i \,\partial_t\, {\CC}_j,}
since the fermions are massive and free, and do not contribute to the
ground-state degeneracy.
This Lagrangian is equivalent to that of a non-relativistic particle
 of mass $1/e^2$ in
a torus of length $2\pi$, threaded by $\kappa$ units of magnetic flux.
Thus,  for integer $\kappa$,
the ground state is a degenerate Landau level of $\kappa$ states. This
generalizes for rational values of $\kappa=q/p$, with
 $(q,p)=1$,  since we may allow multivaluated
wave-functionals. It is enough to focus on the purely topological term
that is obtained by neglecting the Maxwell action (the kinetic energy of
the particle). Then, canonical quantization of the holonomies gives
\eqn\holoq{
e^{i{\CC}_1} \,e^{i{\CC}_2} = e^{2\pi i p/ q}\; e^{i{\CC}_2} \,
e^{i{\CC}_1},}
which can be represented by the $q$-dimensional pair of clock and shift
 matrices. Hence, states are
labelled by the $q$ eigenvalues of, say ${\rm exp}(i\CC_1)$.
Notice that the zero-modes $\CC_i$ are normalized as angular variables  with
period $2\pi$. 

The topological
nature of the Chern--Simons theory allows us to obtain this spectrum by
the general construction of \refs\rwitjons. A basis for the Hilbert
space of the Chern--Simons theory on ${\bf T}^2$ can be obtained by
computing path integrals on the `filled torus' $D\times {\bf S}^1$, with
$D$ a disk carrying an
 insertion of a Wilson line ${\rm exp} \left(iw_1\oint \CA_1\right)$.
 By the usual Chern--Simons `holography', this
basis is in one-to one correspondence with the set of integrable
 representations of the corresponding level-$\kappa$
 WZW model on the boundary of the disk. This restricts the possible values
of $w_1$ to the set
$$
w_1 = 0, 1, \dots, \kappa -1.
$$
 The fact that only one component of the  electric flux
is relevant is seen here by the topological impossibility of `filling in' both
homologically nontrivial circles of the torus at the same time.

A similar construction can be carried out for the non-abelian component
of the index, in terms of the set of integrable representations of
the corresponding $SU(N)$ level-${\bar k}$ WZW model.
Thus, the index can be computed by imposing the ${\bf Z}_N$ modding on
the Hilbert space of the product theory $U(1)_\kappa \times SU(N)_{\bar k}$. 

Notice that there is no shift of the
abelian level by effects of the fermionic measure, since there are
no other fields in the theory that are charged with respect to the diagonal
$U(1)$ subgroup.
Thus, we can freely adjust it depending on the physical definition we adopt for
the abelian level.

 If we wish the $U(N)$ symmetry to act classically, we
define the abelian field 
 $A= {\CA} \cdot {\bf 1}_N$. The periodic $U(1)$-gauge transformations
impose an angular identification, modulo $2\pi$, to the
 holonomy variables $\oint \CA $.
Hence, the effective
abelian level is $\kappa = kN$. However, since the space of ground states
is equivalent to that of the effective Chern--Simons model that arises
after fermion integration, it is more natural to define the $U(N)$ symmetry
referred to this effective bosonic Lagrangian.  In this case, we would
have $\kappa = {\bar k} N$ and the classical theory would be defined with
an extra abelian counterterm with coupling $\delta \kappa = -N^2 /2$. 
Yet another possibility would be to define the abelian level so that the
$U(N)$ symmetry has a simple action  on expectation values of Wilson lines. Since
these are naturally functions of ${\bar k} +N$, one would have a natural
definition $\kappa = N({\bar k} +N)$. 
 
In the particular case $\kappa = {\bar k}\,N$,  
  the purely bosonic low-energy Chern--Simons model is a standard
$U(N)_{\bar k}$ model that can be analyzed directly in terms of the $U(1)^N$
Cartan subalgebra.
 The problem is exactly given by the multi-particle generalization
of \efla\ to a system of $N$ identical bosons. The degeneracy is given by
all the wave functions on $N$ variables  that can be constructed out of
$k$ elementary single-particle `orbitals'. This is
\eqn\unin{
I_{U(N)_{\bar k}} = {(N+{\bar k} -1)! \over N! \,({\bar k} -1)!}. }        
This is 
  the total dimension of the Hilbert space of the product theory $U(1)_{{
\bar k}N} \times SU(N)_{\bar k}$, 
$$
{\bar k}\,N \times {(N+{\bar k} -1)! \over (N-1)! \,{\bar k}!}
$$
 divided by $N^2$, and one can think of the modding by ${\bf Z}_N$ as
dividing by a factor of $N$ for each independent cycle of the torus.
 The general case with asymmetric levels is more
involved. In general, one can argue that the consistency of the ${\bf Z}_N $
modding procedure requires $\kappa= {\bar k}\,N $ (mod $N^2$). Then, a natural
conjecture would be that the index is given by that of the product
$U(1)_{\kappa} \times SU(N)_{\bar k}$ theory, divided by $N^2$. Notice that
the selection rule for $\kappa$ ensures that the result is always an integer.

\newsec{A Formal Digression}

On general grounds, the large-$n$ limit implied in our construction
of ${\overline{U(N)}}_\Theta$ is intimately connected with the definition
of the gauge group for noncommutative theories. In \refs{\rharvey, \rszabo}
 (see also
\refs\rrev) a preliminary analysis is made of this issue. In the traditional
Weyl--Moyal correspondence we consider the space of functions on some space,
in our case ${\bf T}^D$, with some particular smoothness conditions, and the
standard product of functions is deformed into a
$\star$-product. Under appropriate
mathematical conditions, the functions on the noncommutative torus can be
mapped into operators on an infinite-dimensional, separable Hilbert space
$\CH$. The noncommutative gauge  symmetry acts as a subgroup of the
group of unitary transformations on $\CH$, $U(\CH)$. Which subgroup we choose
will determine the gauge-group topology and it is likely also to determine
important non-perturbative properties of the theory.

 In our construction
of ${\overline{U(N)}}_\Theta$ we approximate $\Theta$ by rationals $\Theta_n$
and then, through a Morita transformation, we relate the noncommutative
theories with an ordinary $U(N'_n)$ theory with a certain amount of magnetic
flux, and $N'_n \rightarrow \infty$ as $\Theta_n \rightarrow \Theta$, irrational.
The details of the construction
imply that in this procedure we are keeping the topology of the gauge group
for each intermediate $N'_n$. From this point of view the gauge group looks
like the inductive limit of $U(N)$, i.e. $U(\infty)$, whose homotopy groups
are given by Bott periodicity. A theorem of Palais (c.f. \refs\rpalais)
 implies that
$$
U(\infty) \subset U_1 (\CH) \subset \cdots \subset
 U_p (\CH) \subset \cdots \subset U_{\rm cpt} (\CH),
$$
(where $U \in U_p (\CH)$ iff $U= 1+ \CO$, with $\CO^p$ being trace class, and
$U_{\rm cpt} (\CH)$ are compact operators)
all have the same homotopy type. Hence the proposal in ref. \refs\rharvey\
to consider $U_{\rm cpt} (\CH)$ as the appropriate gauge group arises
as a natural one.

If we were to carry out the naive quantization of the theory with such a gauge
group, it seems clear that the electric and magnetic sectors of the corresponding
gauge theory would `remember' the electric and magnetic sectors of
$U(N)$ for each $N$. One could in principle imagine bigger gauge groups (all
the way to $U(\CH)$, which is contractible). In this case once Gauss' law is
imposed we will lose many of the electric and magnetic sectors of the standard
analysis, or perhaps the new theory would correspond to specific averages of
the standard sectors. In the case of $U(1)$ on ${\bf T}^2_\theta$, the
classical algebra of infinitesimal $U(1)$-gauge transformations is equivalent
to the Fairlie--Fletcher--Zachos algebra, a trigonometric deformation of the
algebra of area-preserving
 diffeomorphisms on ${\bf T}^2$,  $w_\infty ({\bf T}^2)$ (cf. \refs\rzac.)
Once again, there are many choices that could be made to define the quantum
theory with this gauge group. One of them in particular uses the $U(\infty)$
construction above. Other constructions might involve the different definition
of $W_\infty$ starting with $w_\infty$ (see \refs\rcapp\ for details and 
references.) It is clear that the physics depends on the choice of the gauge
group, but at this stage it is not known to what extent.
Addressing this question is likely to be one of the
critical elements in the direct construction of noncommutative gauge
theories for arbitrary values of $\Theta$.

\newsec{Morita Duality and Topological Order in the Quantum Hall Effect}

\noindent

The Morita transformation of the Chern--Simons action \samec\ and, in
particular, the topological coupling \sectres, has an interesting
application to the model of the Fractional Quantum Hall Effect (FQHE)
proposed in \refs\rsuskqh\ (see also \refs{\rbqh, \rram}). Noncommutative
Chern--Simons with rank one was proposed in \refs\rsuskqh\ as a more
refined low-energy description of the FQHE than the usual $U(1)$
Chern--Simons model \refs{\rcsqh, \rbacsus}. Specifically, one considers the
rank-one model
\eqn\csk{
S= {1\over 4\pi \nu} \int_{{\bf R} \times {\bf R}^2_\theta} \left(\CA\wedge
d \CA + {2\over 3} \,\CA \wedge_\star \CA\wedge_\star \CA\right),
}
with a noncommutativity parameter determined
in terms of the electron's
fluid density: $\theta=1/2\pi \rho_e$. The topological
coupling is $1/\nu$, with $\nu$ the FQHE filling fraction. We restrict
for the time being to the Laughlin series $\nu = 1/q$ with odd $q$.
Since noncommutative
 Chern--Simons theories have a very smooth ultraviolet behaviour
 \refs\rcarmelo, we expect the $\theta$-dependence of physical
quantities to be analytic near $\theta=0$, so that the effects
of the electron's `granulariry' reflected in $\theta$ can be
expanded in powers of $\theta$ via the Seiberg--Witten map \refs\rSW.

The real impact of the proposal \csk\ on the physics of the FQHE
should be evaluated on the basis of the induced theory for edge states.
Extracting the boundary dynamics
induced by the bulk action \csk\ is technically non-trivial.
 In fact, it is natural to expect
that, because of the non-locality of the Moyal products, some
boundary action with non-locality on length scales of $\CO(\sqrt{\theta})$
must be added to the bulk Lagrangian in \csk\ (see
\refs{\rgrandi, \rholo} for results in this direction.)
 It is also possible that
the applicability of \csk\ to FQHE phenomena is reduced to the finite
matrix aproximations, such as those studied in \refs{\rbqh,\rram}, which show
encouraging similarities with the microscopic structure of the
relevant wave-function hierarchies.

One way of testing the smooth dependence on $\theta$ using discrete
criteria is to look at Wen's  topological order \refs\rwen, i.e.
the universal degeneracy of Laughlin's ground state on a torus. This
degeneracy is equal to the ordinary Chern--Simons level $1/\nu$.
If the noncommutative model is to give a good description of the FQHE
fluid, we expect that the degeneracy on the torus is correctly accounted
for.

 Working on a
spatial two-torus, the dimensionless noncommutativity parameter is
rational
\eqn\ratt{
\Theta= {1\over n_e},
}
where $n_e$ is the number of electrons in the torus.
Thus, we can use Morita duality to define the noncommutative model
on the torus via the ordinary $U(n_e)$ Chern--Simons theory with
one unit of magnetic flux and level $k' = n_e/\nu$. As in the previous
sections, if we choose the background field and magnetic flux to vanish
for the noncommutative   $U(1)_\Theta$ Chern--Simons model,
the resulting ordinary model has an effectively periodic $U(1)$ sector.
On the other hand, the $SU(n_e)$ sector is twisted with one unit
of 't Hooft's magnetic flux. This means that the non-abelian sector
does not contribute to the degeneracy of ground states, since there are
no $SU(n_e)/{\bf Z}_{n_e}$
 flat connections on the completely twisted torus.

Therefore, the topological order is determined by the abelian factor
with level $k' = n_e/\nu$. As in the previous section, we must present
the abelian level in the effective normalization relevant to the
physics of the zero modes:
\eqn\nomr{
\kappa'_{\rm eff} = {1\over \nu}.}
Thus, the effective abelian level is invariant under Morita transformations
and we recover the expected result for the topological order.

In fact, since the ordinary non-abelian  Chern--Simons model is a topological
theory with a trivial Hilbert space on the  twisted torus (only the vacuum
remains), we can say that the physics of \csk\ on the torus is rigorously
equivalent to that of the ordinary abelian Chern--Simons model with
the same level. Possible non-trival $\theta$-dependence could only arise
when including the Maxwell term in the effective Lagrangian, i.e. the
mixing with higher Landau levels. On the other hand, if the mixing
with higher Landau levels is significant, it is unlikely that the present
field-theoretical degrees of freedom (the `statistical' Chern--Simons gauge
field) will furnish a good description of the dynamics.

Our results in the previous sections indicate that this situation should
also generalize to  non-abelian models that have been proposed as
effective theories of the incompressible Hall fluid with more general
filling fractions \refs\rsuskqh.
Namely, according to \ord,
 a noncommutative $U(N)$ model with   $\Theta =1/n_e$ can be traded by
a twisted ordinary theory with gauge-field momentum modes
 in $U(N) \otimes U(n_e)$
and flat connections living only on the first factor. Since the
effective level, as seen by the flat connections, does not change under
Morita duality, we conclude that the ground-state degeneracy is
independent of $\theta$ also in this generalized situation. 

For example, if a $\nu = p/q$ multi-layer Hall fluid is represented by
a  $U(p)$ noncommutative model at level $q$, the results in the
previous section yield   
a value of the topological order:
\eqn\tono{
{\rm dim}\;\CH({\bf T}_2) = {(p+q -1)! \over p!\,(q-1)!}.}
What is specifically `noncommutative' in this prediction is the particular
`linking' of the global $U(1)$ group and the $SU(p)$ non-abelian part. If
this model is regarded as a single-layer Hall fluid with non-abelian
statistics, the predicted topological order differs in general from other
schemes. For example, an abelian model of type \refs\rfzee\
  for the main Jain sequences, $\nu = p/q$
with 
$q=2ps+1$ and $s$ integer, has enhanced $U(1) \times SU(p)_1$ affine symmetry,
but the topological order is still given by
\eqn\tonoo{
{\rm dim}\;\CH({\bf T}_2)_{\rm Jain} = q = 2ps+1.}
The success of the standard schemes suggests that the noncommutative
non-abelian models are unlikely to describe single-layer fluids along
the main sequences. Therefore, their possible
 applications would be restricted
to multi-layer fluids.  

\subsec{A Comment on Level Normalizations}

\noindent

In the Conformal Field Theory approach to the QHE (see \refs{\rcappelli,
\rfroh} for
a summary,) the topological order
is given by the level of the abelian current algebra of edge excitations.
This is in turn equal to the classical coupling $k$ of the Chern--Simons
Lagrangian. In the noncommutative case, many aspects of the perturbation theory
of the rank-one model \csk\ are similar to the behaviour of non-abelian
Chern--Simons theory. One such instance is the quantum shift $k\rightarrow
k+1$ of the level, analogous to the shift $k\rightarrow k+N$ in ordinary
$SU(N)$ Chern--Simons theory \refs\rshift. To a large extent, the quantum shift
in ordinary Chern--Simons theory is a matter of renormalization prescription,
although many observables, such as expectation values of Wilson lines,
are conveniently written in terms of the shifted level.

The important
point for us is that such a shift does not affect the evaluation of
the dimension of the Hilbert space on the torus, which is still given
in terms of the classical Chern--Simons coupling.
This is rather clear in our computation:  the topological order
is only sensitive to the diagonal $U(1)$, for which there
is no level-shift, since no fields are charged with respect to this
subgroup.

In the context of the finite-matrix models of
\refs\rbqh, the link between the filling fraction and the Chern--Simons
level {\it does} reflect the shift: $k+1 = 1/\nu$. However, these models
contain additional `matter' fields in the fundamental representation
of the $U(N)$ group. Hence, direct comparison of the finite-matrix level
and the one appearing in
 \csk\ may require due attention to renormalization effects
induced when `integrating-out' the matter fields.

\newsec{Conclusions}

\noindent

We have studied some properties of models in
 the dense completion
of rational noncommutative gauge theories, defined
 in finite volume. These models, denoted ${\overline{U(N)
}}_\Theta$,
 are defined
by $\Theta_n \rightarrow \Theta$ limits of rational theories $U(N,\Theta_n)$.

Assuming that these  limiting models exist, they
 are interesting for two reasons. First, one could
try to use them in giving a constructive definition of the
generic noncommutative gauge theory. Even if such a program should fail and
 the  ${\overline{U(N)}}_\Theta$ theory
turned out to be  generically
inequivalent to some other independent definition of the {\it irrational}
$U(N)_\Theta$ theory, the physics of the ${\overline{U(N)}}_\Theta$ is
certainly interesting in itself.

The second interesting aspect of the ${\overline{U(N)}}_\Theta$ models
is that, using Morita duality, we may as well define them as
certain large-$n$ limits of {\it ordinary} gauge theories.  Thus, to
the extent that smooth behaviour in $\Theta$ could have consequences
for the physics of ordinary theories, one would be interested in precisely
 the ${\overline{U(N)}}_\Theta$ models. There may be an interesting
space of large-$n$ limits in ordinary gauge theories that remain
to be explored, and of which the noncommutative theories on tori
are just examples.

The evidence for continuity of the physics
as a function of $\Theta$ is strong for theories with 
an underlying $\CN=4$ supersymmetry.
In models with less supersymmetry, smoothness in $\Theta$ is expected
in perturbation theory, except perhaps at $\Theta=0$. There are exceptional
values of the momenta in perturbation theory for which the rational and
irrational theories behave very differently, but these modes seem to
decouple in the  irrational limit.
At any rate, the most interesting tests of $\Theta$-continuity would
involve non-perturbative physics such as that of
 confining $\CN=1,0$
models.

Using the definition of ${\overline{U(N)}}_\Theta$ in terms of series
of ordinary theories, we have used various expectations about
the dynamics of ordinary confining theories to put constraints on the
$\Theta$-dependence of the ${\overline{U(N)}}_\Theta$ models.
In particular, we find that the Witten index (when properly defined,
so that it gives non-trivial information) of minimal $\CN=1$ models
in three and  four dimensions is smooth under rational approximations.
In the four-dimensional  case, this result depends on the subtle
interplay between the instanton angle and the magnetic flux, exactly
as in the dynamics of oblique confinement.

 A related interesting question is the dependence of the energy on the instanton
angle, a truly non-BPS quantity.
 An estimate of the vacuum energy as a function of $\vartheta$
can be given by a dilute-gas
  fractional-instanton approximation in the series of ordinary theories.
 This corresponds
to the dilute-gas  of ordinary noncommutative instantons after Morita
duality. We find that
 the non-trivial level-crossing phenomena that
ensure $2\pi$-periodicity of physics in ordinary theories translate
into a $2\pi/b_n$-periodicity in the noncommutative theories. In the
$b_n \rightarrow \infty$ limit, the non-abelian contribution to
the  $\vartheta$-dependence becomes
either trivial or discontinuous.

 We have checked that the instanton-induced
functional dependence is smooth in $\Theta$ provided no level-crossing
phenomena takes place. Such is the case when $\CN=4$  supersymmetry is restored
at some high energy scale.
On the other hand, if supersymmetry is not restored, level-crossing
may ocurr, resulting in  strictly $\vartheta$-independent energies
in the $\ubar$ model. Such a result violates $\Theta$-smoothness
when the rational series approximate a {\it rational} number.
Therefore, supersymmetry still acts as a `custodian' of continuity
in $\Theta$.  It would be interesting to know whether there are
other scaling or continuum limits that would guarantee continuity
without necessarily requiring an underlying $\CN=4$ supersymmetric theory
in the ultraviolet. In the non-supersymmetric cases, our 
 treatment of the ultraviolet cutoff is not 
rigorous, in the sense that we assume implicitly a commutativity of
Morita equivalence and continuum limit. It is thus possible that, when
the correct physical questions are asked, continuity of the physics in
$\theta$ is borne out. 

There are still quite a number of open questions; in particular
a viable construction of the quantum theory directly
in the noncommutative setting.  In this case one of the problems
to be solved is the question of the appropriate definition of the
gauge group.  It may turn out that some definitions of the noncommutative
theory are completely disconnected from the naive approximations 
presented in this paper.  A less ambitious problem that should be
tractable is to look again at the UV/IR  
 problem at finite volume.
In theories with rational values of $\Theta$ (whether they are gauge
theories or not) the problem can be expressed in terms of  ordinary
field theories whose fields are matrix valued. In this context the 
standard renormalization of the theory can be carried out without
problems, and in particular one can use standard techniques to define
the Wilsonian effective action (with due attention paid to the
presence of electric and magnetic fluxes.)  The peculiar UV/IR mixing pointed
out in \refs{\ruvir, \rsut}\ would appear as we take combinations of limits where
$\Theta$ becomes irrational, or the volume of the torus goes to
infinity or both.  In any of these cases (see subsection {\it 3.4})
we end up with limits involving large noncommutativity or non-standard
large-$n$ limits.  By looking at rational approximations to the UV/IR
mixing we may gain some understanding of this phenomenon and of its 
physical significance.

Finally we have shown that there are a number of interesting things
that can be learned from Morita duality when applied to the formulation
of the Fractional Hall Effect proposed in \refs\rsuskqh; in paticular we get
an elegant derivation of Wen's topological order \refs\rwen.  Since
Chern--Simons theory is rather soft in the ultraviolet, we do not expect
any problems concerning continuity in the noncommutative parameter.
For these  applications however, this problem is not relevant
since the dimensionless deformation parameter is always rational,
and given by the inverse of the number of electrons in the torus.
A very relevant question is how to extract the boundary dynamics
of edge states induced by the bulk action (8.1).  It would be interesting
to know how to define noncommutative spaces with `boundaries'.
Perhaps in this context the use of the Morita equivalence to an
ordinary gauge theory with gauge group $U(n_e)$ should provide 
useful guidance. 

{\it Note added}.  When this paper was being finished, a paper 
by Z. Guralnik \refs\rzackul\ appeared studying noncommutative
QED using also Morita equivalence and large-$N$ limits of 
ordinary gauge theories.

\vskip1.0cm

\noindent {\bf Acknowledgements}

\vskip 0.1cm

\noindent

We are indebted to Pierre van Baal, Andrea Cappelli,  
 Margarita Garc\'{\i}a P\'erez, C\'esar G\'omez,
 Luca Griguolo, Zack Guralnik,  Jos\'e Labastida,
 Esperanza L\'opez,   Eliezer
Rabinovici, Domenico Seminara, Albert Schwarz and Paolo Valtancoli
 for  useful discussions. We also thank Angel Paredes for finding
an error in a previous version of subsection {\it 6.3}.
  L.A.-G. would like to
thank the hospitality of the Humboldt Universtity at
Berlin, where part of this work was done, and in particular
Dieter L\"ust.

\listrefs

\bye